\documentclass[aps,prb,reprint,superscriptaddress]{revtex4-2}

\usepackage[dvipdfmx]{graphicx}
\usepackage[T1]{fontenc}
\usepackage{textcomp}
\usepackage{lmodern}
\usepackage{amsmath,amssymb,mathtools}
\usepackage{bm}
\usepackage{graphicx}
\usepackage{hyperref}
\usepackage{microtype}
\usepackage{siunitx}
\usepackage{physics} 
\usepackage{comment}

\hypersetup{
	colorlinks=true,
	linkcolor=blue,
	citecolor=blue,
	urlcolor=blue, 
	breaklinks=true
}

\begin{document}

\title{A cascade of orbital-field-driven phases with a Lifshitz-type transition to unidirectional spiral order}
\author{Hiroki Nakai}
\email{hiroki.nakai@utoronto.ca}
\affiliation{Department of Physics, University of Toronto, 60 St. George Street, Toronto, Ontario, M5S 1A7, Canada}
\author{Hae-Young Kee}
\email{hy.kee@utoronto.ca}
\affiliation{Department of Physics, University of Toronto, 60 St. George Street, Toronto, Ontario, M5S 1A7, Canada}
\affiliation{Canadian Institute for Advanced Research, CIFAR Program in Quantum Materials, Toronto, Ontario, M5G 1M1, Canada}

\date{\today}

\begin{abstract}
	Frustration in magnetic systems gives rise to unconventional orders and exotic spin liquids. Here, we explore a new type of frustration arising from an intrinsic bond-dependent \emph{orbital-field} that couples orbital degrees of freedom to spin correlations.
	Introducing this orbital-field interaction into the antiferromagnetic Heisenberg model on the honeycomb lattice, we show that 
	local spin-orbital constraints generate a subextensive degeneracy, lifted by order-by-disorder to select distinct collinear magnets in a cascade of phases as the orbital field increases from a para-orbital liquid. 
	At larger fields, the selected phase gives way to an incommensurate unidirectional spiral that spontaneously breaks chiral symmetry.
	This transition is driven by the coupling of a Goldstone mode to gapped dynamical spin-orbital fluctuations. The spin stiffness becomes negative along a particular direction, while the spiral wavevector grows as the square root of the distance from the transition, signaling a Lifshitz-type transition.
	We present a microscopic route to realizing this orbital field and discuss candidate materials.
\end{abstract}

\maketitle

\section*{Introduction}
Nature often favors a ground state that satisfies all local interactions. When such a selection is hindered, frustration arises. In magnetic systems, frustration plays a central role in the emergence of exotic phases of matter, including quantum spin liquids and unconventional magnetic orders~\cite{Lacroix2011}. Magnetic frustration can arise from two distinct mechanisms: geometric frustration, exemplified by Ising antiferromagnets on triangular lattices~\cite{Wannier1950, Stephenson1964, Moessner2001}, and interaction-driven frustration, as found in the bond-dependent Kitaev interaction~\cite{Kitaev2006, Jackeli2009, Chaloupka2010, Rau2014}.

Beyond these conventional mechanisms, the interplay between spin and orbital degrees of freedom provides another route to frustration~\cite{Kugel1982, Tokura2000, Khaliullin2005, Streltsov2017}. A paradigmatic example is the Kugel-Khomskii model~\cite{Kugel1973}, in which spin and orbital degrees of freedom are coupled through superexchange interactions, leading to a rich variety of competing spin and orbital orders~\cite{Feiner1997, Khaliullin2000, Vernay2004, Normand2008, Jackeli2008, Brzezicki2012, Savary2021, Churchill2025}. The directional nature of orbital interactions can also give rise to compass models~\cite{Nussinov2015}, where different orbital components are coupled along different bond directions. Such bond-dependent interactions introduce intrinsic frustration and can stabilize unconventional orbital orders and highly degenerate ground states~\cite{Khomskii2003, Mishra2004, Biskup2005, Wu2008, Nasu2008}.

Here, we explore a distinct type of frustration arising from a bond-dependent spin-orbital interaction that is linear in orbital operators. We refer to this interaction as the \emph{orbital-field}, which couples bond-dependent orbital components directly to spin correlations. Unlike conventional orbital exchange, the orbital field itself generates frustration through its bond dependence, even in the absence of bilinear orbital interactions.
While a conventional field, such as a Zeeman field, favors polarization, the intrinsic orbital field generates magnetic frustration because spin correlations on different bonds couple differently to the same orbital. This prevents the simultaneous optimization of orbital configurations and spin correlations, imposing local spin-orbital constraints.

We investigate this mechanism using a minimal spin-orbital model on the honeycomb lattice, combining nearest-neighbor antiferromagnetic (AFM) Heisenberg exchange with the orbital-field interaction. We find that this coupling generates a cascade of competing phases, dimensional reduction, and a Lifshitz-type transition to unidirectional spiral order. Our findings uncover a distinct source of frustration, revealing a new route to unconventional magnetic orders in spin-orbital systems.

\section*{Results}

\subsection*{Orbital-field coupled to spin correlation}

\begin{figure}[t]
	\centering
	\includegraphics[width=8cm]{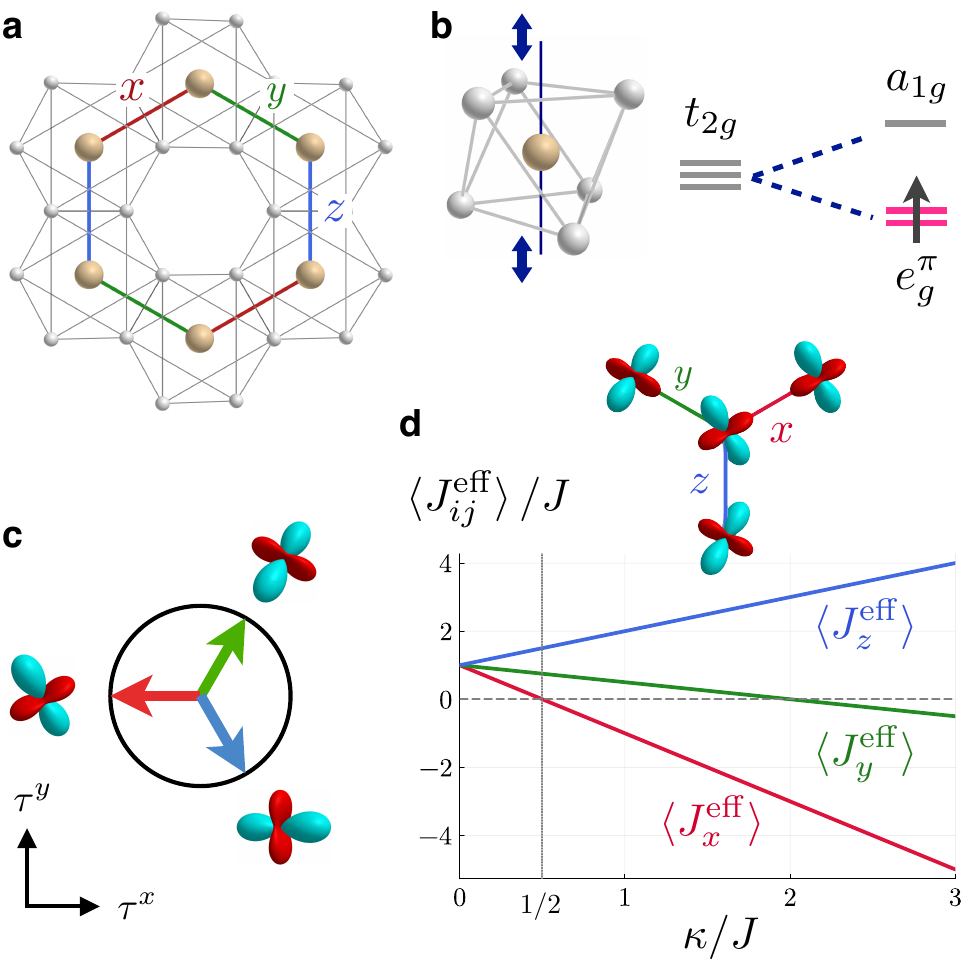}
	\caption{
		\textbf{Spin-orbital model on the honeycomb lattice.}
		\textbf{a} Honeycomb lattice with the three nearest-neighbor bonds $\gamma=x,y,z$.
		\textbf{b} Trigonal crystal-field splitting of the $t_{2g}$ orbitals into a lower $e_g^\pi$ doublet and an upper $a_{1g}$ singlet, with one electron occupying the $e_g^\pi$ doublet in the $d^1$ configuration.  
		\textbf{c} Orbital pseudospins in the $\tau^x$-$\tau^y$ plane and the corresponding real orbital wave functions. 
		The red, green, and blue arrows represent the three orbital pseudospin states satisfying $\langle\tilde{\tau}^\gamma\rangle=-1/2$ for $\gamma=x,y,z$, respectively. 
		\textbf{d} Orbital dependence of the effective spin exchange $\langle J_{ij}^{\rm eff}\rangle$ for a representative orbital configuration shown schematically.
	}
	\label{fig:model}
\end{figure}

The orbital-field interaction takes the following form:
\begin{equation}\label{eq:Ham_kappa}
\mathcal{H}_{\rm orb} =2\kappa \sum_{\langle i,j \rangle} (\tilde{\tau}_i^\gamma +\tilde{\tau}_j^\gamma)\, \vb*{S}_i\cdot\vb*{S}_j. 
\end{equation}
Here, the bond-dependent orbital operators are 
\begin{equation}\label{eq:bond_dependent_tau}
\tilde{\tau}_i^\gamma = \cos\phi_\gamma\,\tau_i^x +\sin\phi_\gamma\,\tau_i^y,
\end{equation}
where $\tau_i^\mu$ $(\mu=x,y,z)$ is an orbital pseudospin-$1/2$, with $\phi_x=0$, $\phi_y=-2\pi/3$, and $\phi_z=2\pi/3$ for the three nearest-neighbor bond types $\gamma=x,y,z$ on the honeycomb lattice (Fig.~\ref{fig:model}a).
This interaction couples $\tilde{\tau}^\gamma$ to spin exchange interactions, such that orbital configurations modulate the spin exchange while the resulting spin correlations generate intrinsic orbital fields. 
As discussed below, a microscopic realization of this model can be found in $t_{2g}$ orbital systems under a trigonal crystal field, where conventional intersite orbital exchange can vanish while the intrinsic orbital field remains finite. 

The pseudospin $\vb*{\tau}$ represents an orbital doublet, for which we consider the $e_g^\pi$ doublet arising from the splitting of the $t_{2g}$ orbitals under a trigonal crystal field~\cite{Chern2010, Kugel2015, Nakai2026}, as shown in Fig.~\ref{fig:model}b. 
The $e_g^\pi$ doublet corresponds to the effective orbital angular momentum states $l_{\rm eff}^z=\pm 1$, which transform as $\ket*{\pm 1} \to -\ket*{\mp 1}$ under time reversal [see Eq.~(S63) in Supplementary Materials (SM)]. 
Thus, $\tau^x$ and $\tau^y$ represent the orbital quadrupoles, while $\tau^z$ represents an orbital dipole.
Figure~\ref{fig:model}c shows three orbital states with pseudospins separated by angles of $2\pi/3$ in the $\tau^x$-$\tau^y$ plane, together with their corresponding real orbital wave functions. 
These states satisfy $\langle\tilde{\tau}^\gamma\rangle=-1/2$ for $\gamma=x,y,z$, respectively. 

\begin{figure*}[t]
	\centering
	\includegraphics[width=17.6cm]{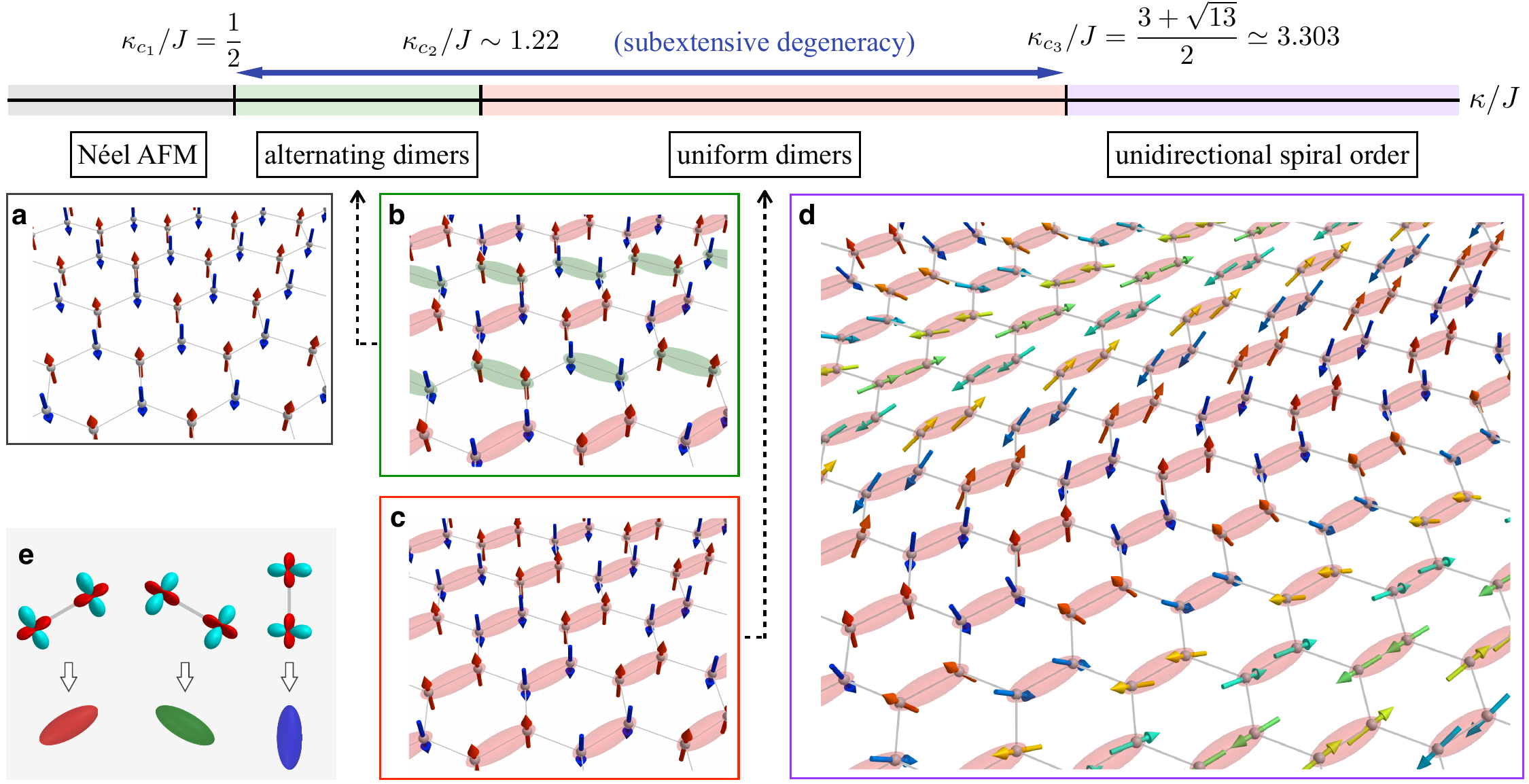}
	\caption{
		\textbf{A cascade of orbital-field-driven phases.}
		Semiclassical phase diagram as a function of $\kappa/J$, obtained from site-product variational calculations and LFW theory. 
		\textbf{a-d} Schematic spin-orbital configurations of the N\'eel AFM (\textbf{a}), alternating-dimer (\textbf{b}), uniform-dimer (\textbf{c}), and unidirectional spiral (\textbf{d}) phases.
		Arrows denote spins, and colored ovals indicate orbital configurations associated with FM spin pairs. 
		In the N\'eel phase, no dimers are present, and the orbital degrees of freedom remain unconstrained, corresponding to a para-orbital liquid. 
		\textbf{e} Correspondence between orbital configurations and colored ovals, with red, green, and blue representing $x$-, $y$-, and $z$-bonds, respectively.
		Each dimer on a $\gamma$-bond fixes the orbital pseudospins at its ends to satisfy $\ev*{\tilde{\tau}_i^\gamma}=-1/2$. 
		The region labeled ``subextensive degeneracy'' denotes the ground-state manifold before the LFW zero-point correction is included.
	}
	\label{fig:phase_diagram}
\end{figure*}

\subsection*{Semiclassical phase diagram}

To investigate how the orbital-field generates frustration, 
we consider the AFM Heisenberg model supplemented by the orbital-field interaction on honeycomb lattice: 
\begin{equation}\label{eq:Ham}
\mathcal{H} = J \sum_{\langle i,j\rangle} \vb*{S}_i\cdot\vb*{S}_j +\mathcal{H}_{\rm orb} \equiv \sum_{\langle i,j\rangle} J_{ij}^{\rm eff} \vb*{S}_i\cdot\vb*{S}_j, 
\end{equation}
where $\vb*{S}_i$ denotes the spin-$S$ moment at site $i$, and $ J_{ij}^{\rm eff} \equiv J+2\kappa(\tilde{\tau}_i^\gamma+\tilde{\tau}_j^\gamma)$.
Throughout this work, we consider $S=1/2$ and focus on the regime $J>0$ and $\kappa>0$.

To understand how the orbital field generates frustration at the bond level, we consider a fixed orbital configuration.
For a given orbital configuration, the orbital-dependent effective exchange is given by $\langle J_{ij}^{\rm eff} \rangle =J+2\kappa(\langle \tilde{\tau}_i^\gamma \rangle +\langle \tilde{\tau}_j^\gamma \rangle)$.
Since $\sum_{\gamma=x,y,z}\tilde{\tau}_i^\gamma=0$, the orbital-field interaction cannot modify the exchanges along all three bond directions uniformly. Enhancing the exchange along some bonds necessarily weakens it along others, as illustrated by the orbital-field strength $\kappa$ dependence of $\langle J_\gamma^{\rm eff}\rangle$ for a representative orbital configuration in Fig.~\ref{fig:model}d.
For $\kappa<J/2$, all three $\langle J_\gamma^{\rm eff}\rangle$ remain AFM, whereas for $\kappa>J/2$, orbital modulation can drive one of them ferromagnetic (FM) while the other two remain AFM. 

To determine which spin-orbital configurations are favored by this competition, we investigate the ground states of the model using site-product variational states that allow for local spin-orbital entanglement (see Methods).
The resulting phase diagram is summarized in Fig.~\ref{fig:phase_diagram}.
For $\kappa/J<1/2$, the spin forms a N\'eel state coexisting with a para-orbital liquid, in which the orbital degrees of freedom remain unconstrained (see Fig.~\ref{fig:phase_diagram}a). 
At $\kappa_{c_1}/J=1/2$, the system leaves the N\'eel phase as one of the three effective exchanges changes sign. 

Beyond this point for $ 1/2 < \kappa/J < 3.303$, the orbital configurations give rise to a one-FM and two-AFM pattern around each site, providing the local constraints. 
Representing each FM bond as a dimer (indicated by colored ovals)  maps the spin-orbital configurations onto dimer coverings of the honeycomb lattice, with each dimer fixing the orbital states at its ends, as shown in Fig.~\ref{fig:phase_diagram}e. 
This intermediate-$\kappa$ regime hosts a subextensively degenerate manifold of collinear spin states comprising parallel zigzag chains, each covered by dimers occupying one of the two bond types along the chain, including the alternating and uniform dimer configurations shown in Figs.~\ref{fig:phase_diagram}b,c, respectively. 
We will discuss the origin of the subextensive degeneracy below before turning to the large orbital-field regime, where a unidirectional chiral spiral order (see Fig.~\ref{fig:phase_diagram}d) emerges through a Lifshitz-type transition.

\begin{figure*}[t]
	\centering
	\includegraphics[width=16.0cm]{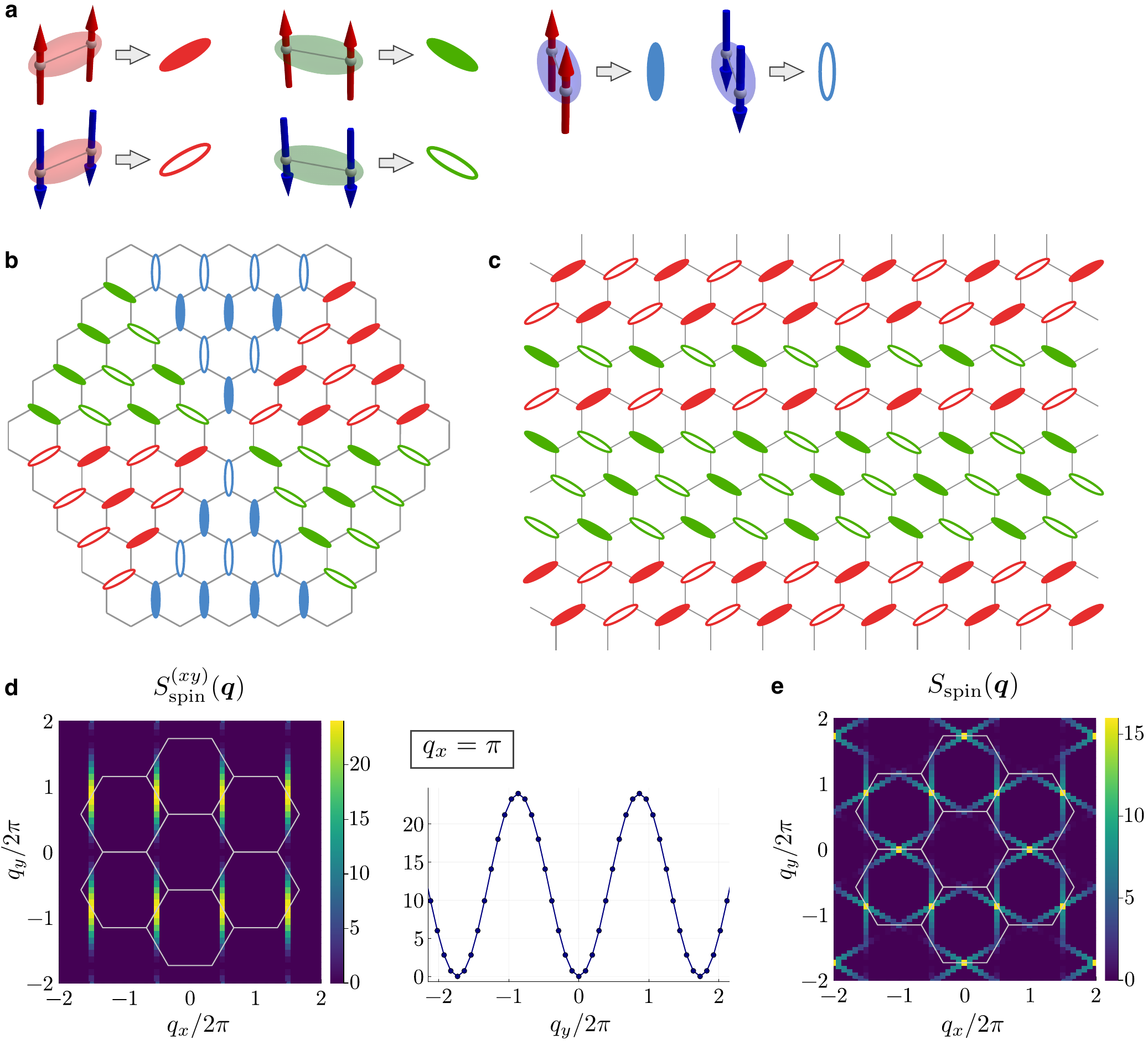}
	\caption{
		\textbf{Dimer configurations with subextensive degeneracy.} 
		\textbf{a} Mapping of the spin-orbital configurations onto a dimer representation.
		FM spin pairs, indicated by spin arrows and colored ovals, are mapped onto filled and open dimers for up-up and down-down spin pairs, respectively. 
		The oval colors and corresponding orbitals follow the convention in Fig.~\ref{fig:phase_diagram}\textbf{e}. 
		\textbf{b} Dimer configuration with $n_d=0$ on a hexagonal plaquette. 
		\textbf{c} Representative ground state consisting of parallel $xy$-chains with $x$- or $y$-bond dimers.
		\textbf{d} Spin structure factor $S_{\rm spin}^{(xy)}(\vb*{q})$ for the $xy$-chain ground-state manifold at $L=12$, with a line cut at $q_x=\pi$ (right).
		\textbf{e} Spin structure factor $S_{\rm spin}(\vb*{q})$ averaged over the full ground-state manifold at $L=12$.
	}
	\label{fig:dimer}
\end{figure*}

\subsection*{Subextensive degeneracy and dimensional reduction}

We now examine the origin of the subextensive degeneracy found at intermediate-$\kappa$ regime ($0.5 < \kappa/J < 3.303)$.
The structure of the allowed configurations becomes transparent in the dimer representation introduced above.
Each dimer corresponds to an FM bond connecting parallel spins. 
For a site belonging to a dimer on a $\gamma$-bond, the one-FM and two-AFM bond configuration, together with $\sum_{\gamma=x,y,z}\tilde{\tau}_i^\gamma=0$, reduces the local orbital-field energy to $\kappa \tilde{\tau}_i^\gamma$. 
Minimizing this energy fixes the orbital pseudospins at both ends of the dimer to satisfy $\ev*{\tilde{\tau}_i^\gamma}=-1/2$, as shown in Fig.~\ref{fig:phase_diagram}e. 
We further distinguish up-up and down-down spin pairs by filled and open ovals, respectively, as shown in Fig.~\ref{fig:dimer}a. 
Spins on neighboring dimers must be antiparallel across the remaining AFM bonds, a requirement we refer to as the interdimer AFM condition.
The ground-state condition is therefore equivalent to a complete dimer covering satisfying this condition.

To characterize these coverings, let $n_d$ denote the number of dimers on the edges of a hexagon ($0\leq n_d\leq3$). 
The interdimer AFM condition excludes $n_d=1$ and $3$, leaving $n_d=0$ and $2$ as locally allowed configurations.
For $n_d=0$, specifying the configuration on a single hexagon uniquely determines the dimer covering throughout the lattice (see Fig.~\ref{fig:dimer}b), yielding at most $O(L^2)$ configurations for a system of linear size $L$.

The much larger family of ground-state configurations, however, has $n_d=2$ on every hexagon, forming parallel zigzag chains with dimers occupying one of the two bond types along each chain. 
Figure~\ref{fig:dimer}c shows a representative configuration of zigzag chains formed by alternating $x$- and $y$-bonds, referred to as $xy$-chains. 
Along each chain, dimers occupy either all $x$-bonds or all $y$-bonds, providing two possible dimer patterns. 
The spins exhibit a repeating up-up-down-down pattern, and the orbitals are uniformly aligned within each chain. 
The freedom to choose dimer patterns across chains yields a subextensive degeneracy, $W_{\rm GS}\sim2^L$ (see SM for details of the configurations and their counting). 
Equivalent configurations with $yz$- and $zx$-chains follow from $C_3$ rotations. 
This chain structure gives rise to long-range spin and orbital correlations along the chains, but only short-range correlations between them. 

The one-dimensional correlations are also reflected in the spin structure factor, 
\begin{equation}\label{eq:spin_structure_factor}
S_{\rm spin}(\vb*{q}) = \frac{1}{N} \sum_{i,j} \langle \vb*{S}_i\cdot\vb*{S}_j \rangle e^{-i\vb*{q}\cdot(\vb*{r}_i-\vb*{r}_j)}.
\end{equation}
For the $xy$-chain ground states, $S_{\rm spin}^{(xy)}(\vb*{q})$ exhibits spectral weight confined to lines $q_x=(2m+1)\pi$ ($m\in\mathbb{Z}$), as shown in Fig.~\ref{fig:dimer}d. 
The intensity along these lines varies continuously with $q_y$, reflecting short-range transverse correlations. 
Figure~\ref{fig:dimer}e shows the spin structure factor averaged over the full ground state manifold.
Together, the subextensive degeneracy and one-dimensional correlations demonstrate the frustration-induced dimensional reduction.

\begin{figure}[t]
	\centering
	\includegraphics[width=8cm]{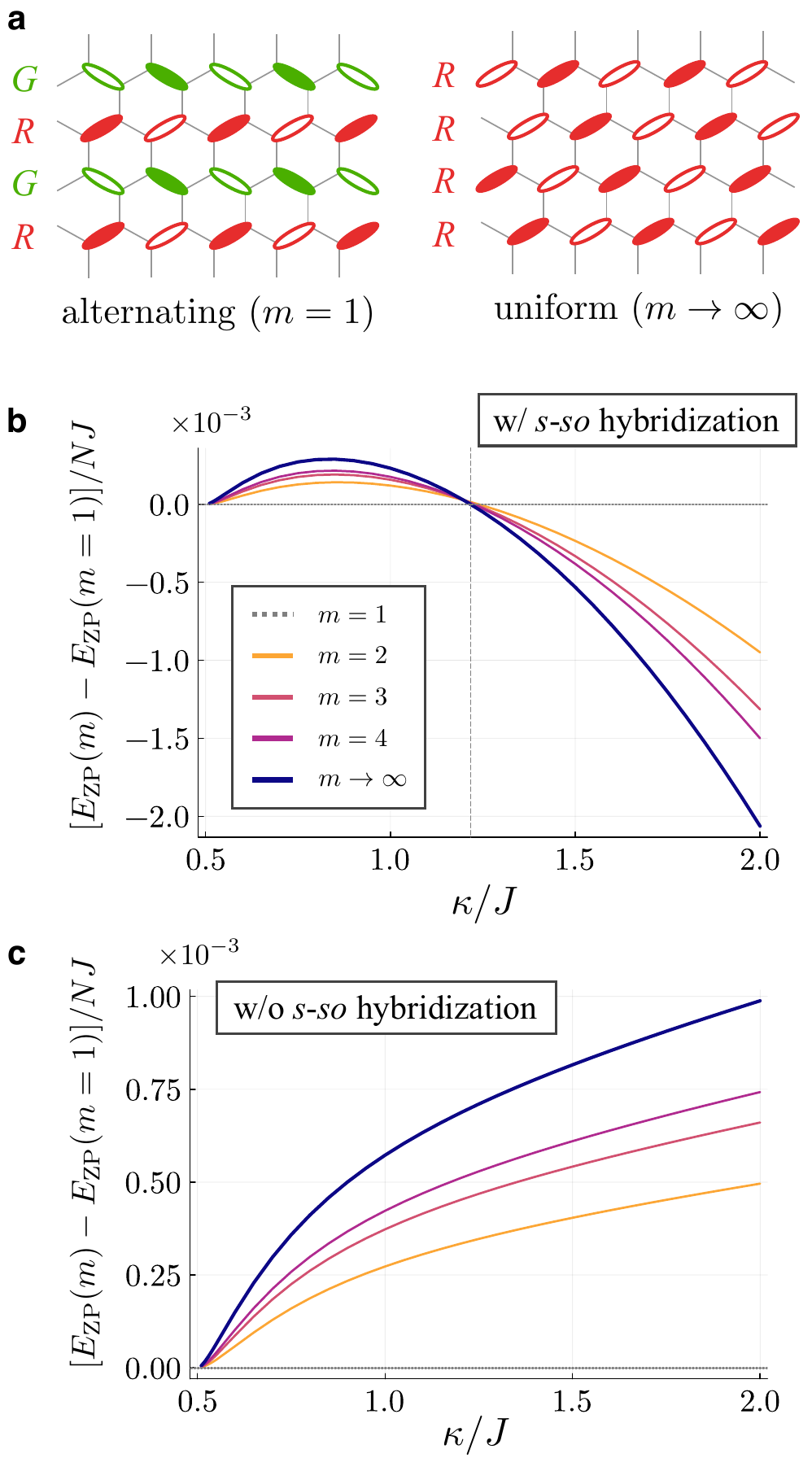}
	\caption{
		\textbf{Quantum selection of dimer configurations.}
		Periodic dimer configurations are denoted by $R^mG^mR^mG^m\cdots$, where $R$ and $G$ represent zigzag chains with dimers occupying the $x$- and $y$-bonds, respectively, and $m$ denotes the number of consecutive chains with the same dimer pattern. 
		\textbf{a} Alternating ($m=1$) and uniform ($m\to\infty$) dimer configurations. 
		\textbf{b} Relative zero-point energy per site, $[E_{\rm ZP}(m)-E_{\rm ZP}(m=1)]/NJ$, for $m=2$, $3$, $4$, and $m\to\infty$, as a function of $\kappa/J$, including the hybridization term between the spin ($s$) and spin-orbital ($so$) modes.
		\textbf{c} Same as \textbf{b}, with the hybridization term switched off.
	}
	\label{fig:ZP}
\end{figure}

\subsection*{Quantum selection of spin-orbital orders}
A natural next question is whether the subextensive degeneracy found in the classical model persists under quantum fluctuations.
We investigate how quantum fluctuations lift the subextensive degeneracy using linear flavor-wave (LFW) theory~\cite{Joshi1999}, which yields the zero-point energies and excitation spectra of the dimer configurations.
At each site, the three states orthogonal to the local spin-orbital state describe spin, orbital, and spin-orbital fluctuations.
Expanding the Hamiltonian to quadratic order yields three corresponding sectors. 
The orbital mode is dispersionless and decoupled from the other two sectors, whereas the spin and spin-orbital modes hybridize through terms such as $(\tilde{\tau}_i^\gamma \vb*{S}_i)\cdot\vb*{S}_j$ in the $\kappa$ term. 
The explicit quadratic Hamiltonian [Eq.~\eqref{eq:LFW_Ham}] is given in Methods.

To compare zero-point energies among different dimer configurations, we consider periodic dimer configurations $R^mG^mR^mG^m\cdots$, where $R$ and $G$ denote zigzag chains with $x$- and $y$-bond dimers, respectively, and $m$ is the number of consecutive chains of each type. 
The two limiting cases are the alternating configuration ($m=1$) and the uniform configuration ($m\to\infty$), as shown in Fig.~\ref{fig:ZP}a.
In addition to these two limits, we examine several intermediate periodic configurations with $m=2$, $3$, and $4$.
Figure~\ref{fig:ZP}b shows their zero-point energies relative to that of $m=1$ as functions of $\kappa/J$. 

At the boundary with the N\'eel phase, $\kappa_{c_1}/J=1/2$, all the dimer configurations have the same zero-point energy.
For $1/2<\kappa/J\lesssim1.22$, the relative energy increases with $m$, selecting the alternating configuration ($m=1$).
The configurations become degenerate again at $\kappa_{c_2}/J\simeq1.22$, beyond which the energetic ordering reverses and the uniform configuration ($m\to\infty$) is selected.

To identify the origin of this transition, we focus on the hybridization term between the spin and spin-orbital modes (w/ $s$-$so$ hybridization in Fig.~\ref{fig:ZP}b), which favors the uniform dimer configuration. In contrast, when this term is switched off (w/o $s$-$so$ hybridization in Fig.~\ref{fig:ZP}c), the alternating configuration remains favored for arbitrarily large $\kappa/J$. This demonstrates that the hybridization term drives the reversal of quantum selection.

\begin{figure*}[t]
	\centering
	\includegraphics[width=16cm]{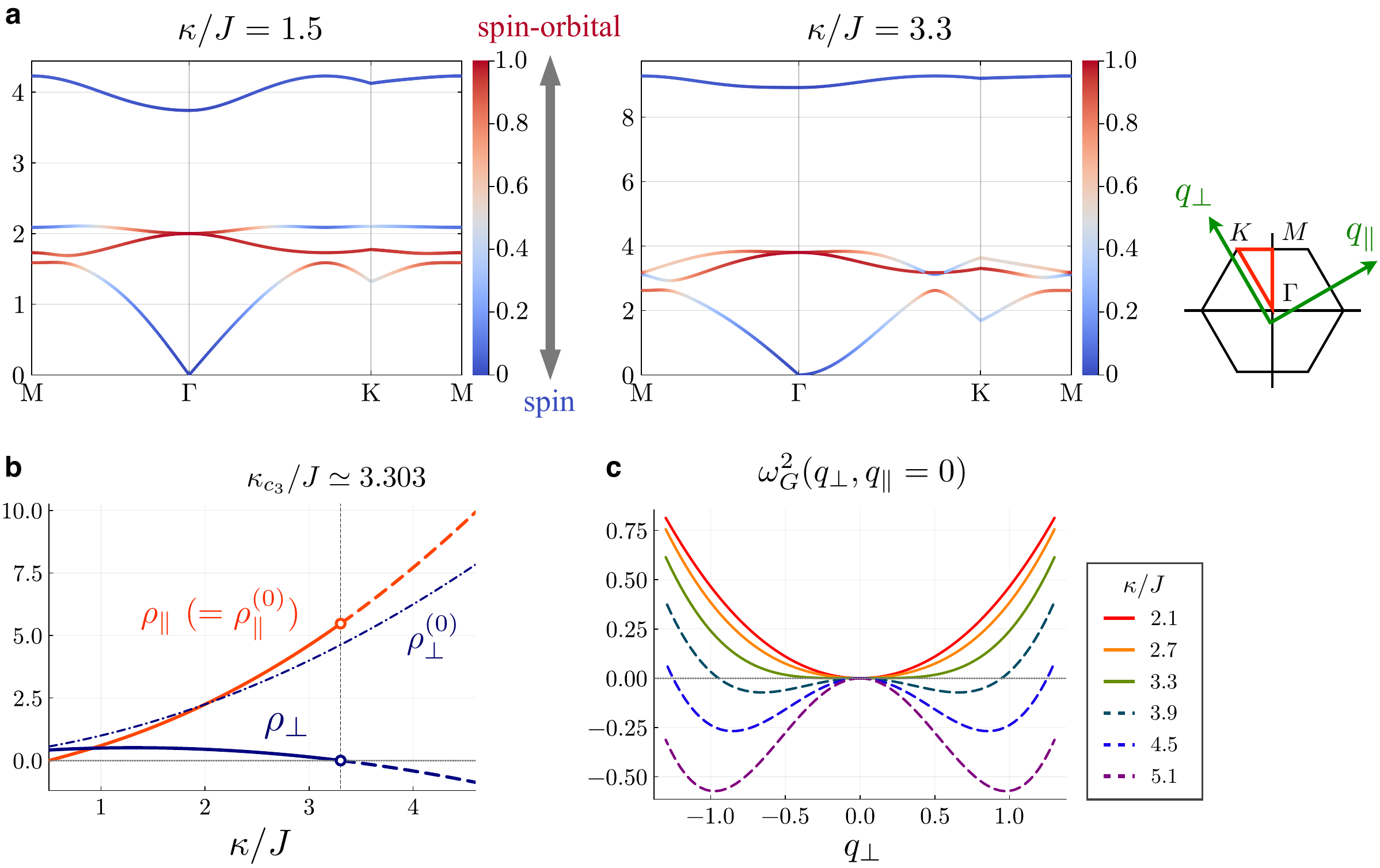}
	\caption{
		\textbf{Goldstone-mode instability induced by spin-orbital fluctuations.}
		\textbf{a} LFW spectra of the uniform-dimer configuration at $\kappa/J=1.5$ and $3.3$, with the spin and spin-orbital weights indicated in blue and red, respectively. 
		The orbital mode is omitted for clarity. 
		\textbf{b} Spin stiffnesses $\rho_\parallel$ and $\rho_\perp$, defined for momenta parallel and perpendicular to the dimers, respectively. The corresponding stiffnesses $\rho_\parallel^{(0)}$ and $\rho_\perp^{(0)}$, obtained with the hybridization between the spin and spin-orbital modes switched off, are shown for comparison. 
		\textbf{c} Evolution of $\omega_G^2$ along the $\Gamma$-K direction across $\kappa_{c_3}$.
		In \textbf{b} and \textbf{c}, dashed lines indicate formal continuations into the unstable regime $\kappa>\kappa_{c_3}$.
	}
	\label{fig:flavor-wave}
\end{figure*}

\subsection*{Goldstone-mode instability from spin-orbital fluctuations and Lifshitz-type transition}

The hybridization between the spin and spin-orbital modes also affects the low-energy excitations. 
Figure~\ref{fig:flavor-wave}a shows the LFW spectra of the uniform-dimer configuration at $\kappa/J=1.5$ and $3.3$, with the spin and spin-orbital weights of each mode. 
The lowest branch is the gapless Goldstone mode associated with spontaneous breaking of the global spin-rotational symmetry. 
While the Goldstone mode exhibits a linear dispersion in both directions at $\kappa/J=1.5$, it remains linear along $\Gamma$-M but becomes nearly quadratic along $\Gamma$-K at $\kappa/J=3.3$, revealing strongly anisotropic softening. 

We characterize this softening through the long-wavelength expansion
\begin{equation}\label{eq:goldstone_stiffness}
\omega_G^2(\vb*{q}) = \rho_\parallel q_\parallel^2+ \rho_\perp q_\perp^2+ \beta_\perp q_\perp^4 +\cdots,
\end{equation}
where $q_\parallel$ and $q_\perp$ denote the momentum components parallel and perpendicular to the dimers, respectively, with $q_\perp$ along $\Gamma$-K.
As shown in Fig.~\ref{fig:flavor-wave}b, $\rho_\parallel$ remains positive, whereas $\rho_\perp$ decreases continuously and vanishes at $\kappa_{c_3}/J=(3+\sqrt{13})/2\simeq3.303$ (see SM for the derivation). 

To understand the stiffness reduction, we consider a low-energy theory of the coupled spin and spin-orbital modes. 
Let $\theta$ denote the Goldstone coordinate and $\phi$ the gapped spin-orbital fluctuation. 
The hybridization term vanishes at $\Gamma$ and grows linearly with $q_\perp$, as derived from the LFW Hamiltonian (see SM for details). 
The resulting low-energy Lagrangian takes the form
\begin{equation}\label{eq:low_energy_lagrangian}
\begin{split}
\mathcal{L}
&= \frac{1}{2}(\partial_t\theta)^2 -\frac{\rho_\perp^{(0)}}{2}(\partial_\perp\theta)^2 -\frac{\beta_\perp^{(0)}}{2}(\partial_\perp^2\theta)^2 \\
&\quad +\frac{1}{2}(\partial_t\phi)^2 -\frac{\Delta_{so}^2}{2}\phi^2 -\gamma\phi\,\partial_\perp\theta +\cdots ,
\end{split}
\end{equation}
where $\Delta_{so}$ is the gap of the spin-orbital mode and $\gamma$ denotes its derivative coupling to the Goldstone mode.

At energies well below $\Delta_{so}$, the spin-orbital field follows the slowly varying Goldstone mode adiabatically.
To leading order in this low-energy regime, the spin-orbital field is given by
\begin{equation}\label{eq:induced_so_fluctuation}
\phi \simeq -\frac{\gamma}{\Delta_{so}^2}\partial_\perp\theta, 
\end{equation}
indicating that slow spatial variations of the spin orientation induce spin-orbital fluctuations, allowing the local orbital configuration to readjust together with the spin and thereby reducing the energy cost of the modulation. 

Integrating out $\phi$ yields an effective theory for the Goldstone mode,
\begin{equation}\label{eq:effective_goldstone_lagrangian}
\mathcal{L}_{\rm eff}
= \frac{1}{2}(\partial_t\theta)^2 -\frac{\rho_\perp^{\rm eff}}{2}(\partial_\perp\theta)^2 -\frac{\beta_\perp^{\rm eff}}{2}(\partial_\perp^2\theta)^2 +\cdots ,
\end{equation}
with the effective spin stiffness in the perpendicular direction given by
\begin{equation}\label{eq:effective_stiffness}
\rho_\perp^{\rm eff} = \rho_\perp^{(0)} -\frac{\gamma^2}{\Delta_{so}^2}.
\end{equation}
The negative correction accounts for the selective softening of the Goldstone mode induced by the hybridization term in the LFW calculation, as shown in Fig.~\ref{fig:flavor-wave}b.

At $\kappa_{c_3}/J = 3.303$, the leading spatial term along the soft direction is $(\partial_\perp^2\theta)^2$. 
The Gaussian low-energy theory acquires an emergent symmetry, $\theta(\vb*{r},t) \rightarrow \theta(\vb*{r},t)+a+b r_\perp$, under which a uniform gradient perpendicular to the dimers carries no energy cost at quadratic order. 
This symmetry is absent from the microscopic Hamiltonian but emerges at the critical point. 

Beyond $\kappa_{c_3}$, the negative quadratic-gradient term competes with the positive higher-gradient term, shifting the instability continuously from $\Gamma$ to finite momentum. 
This behavior is characteristic of an anisotropic Lifshitz-type instability~\cite{Hornreich1975, Redner1977}, in which the stiffness vanishes along one direction while remaining finite along the orthogonal direction. 

Figure~\ref{fig:flavor-wave}c shows how the instability develops in the LFW analysis through the evolution of $\omega_G^2$ along $\Gamma$-K. 
At $\kappa_{c_3}$, the quadratic term vanishes while $\beta_\perp$ remains positive, yielding $\omega_G^2\propto q_\perp^4$ and a quadratic Goldstone dispersion. 
Beyond $\kappa_c$, the uniform-dimer configuration is dynamically unstable; nevertheless, a formal continuation of $\omega_G^2$ reveals how the instability develops. 
The curvature at $\Gamma$ becomes negative,  shifting the minimum continuously to finite momentum and signaling the onset of incommensurate order. 

\begin{figure*}[t]
	\centering
	\includegraphics[width=17.6cm]{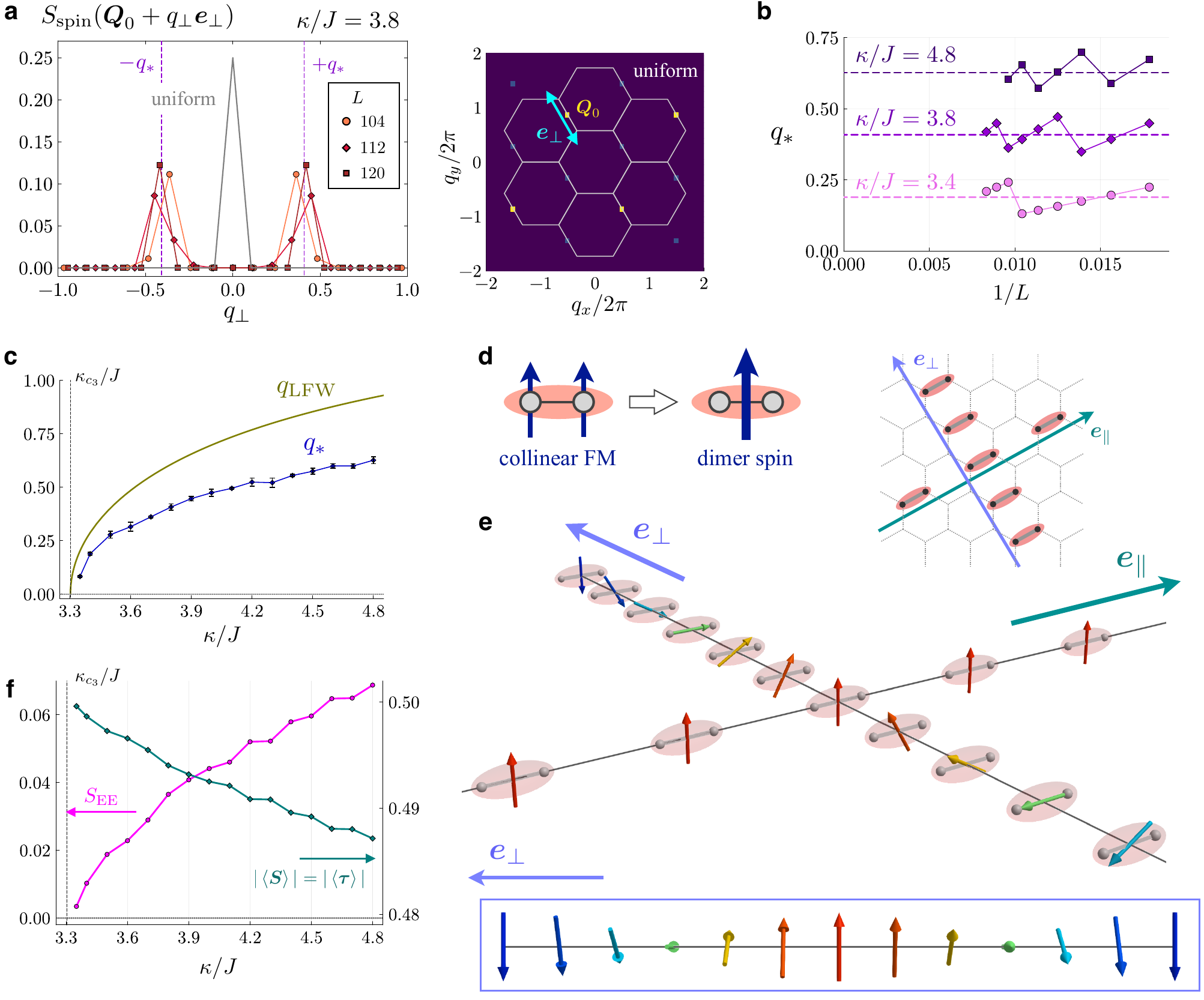}
	\caption{
		\textbf{Unidirectional incommensurate spiral magnet with spin-orbital entanglement.}
		\textbf{a} Spin structure factors at $\kappa/J=3.8$. 
		Left: $S_{\rm spin}(\vb*{Q}_0+q_\perp\vb*{e}_\perp)$ for variationally optimized states at $L=104$, $112$, and $120$, compared with the uniform-dimer configuration at $L=120$.
		Right: $S_{\rm spin}(\vb*{q})$ of the uniform-dimer configuration at $L=12$, showing the ordering wave vector $\vb*{Q}_0$ and modulation direction $\vb*{e}_\perp$.
		\textbf{b} Finite-size dependence of the structure-factor peak position $q_\perp$ and the estimated $q_*$ for $\kappa/J=3.4$, $3.8$, and $4.8$. 
		\textbf{c} Modulation wave vector $q_*$ from variational calculation and $q_{\rm LFW}$ from the minimum of $\omega_G^2$, as functions of $\kappa/J$. 
		For $q_*$, the symbols denote the midpoints of the upper and lower bounds estimated from the finite-size analysis, while the vertical bars indicate the corresponding bounds.
		\textbf{d} Schematic dimer spin representing two parallel spins on an FM bond. 
		\textbf{e} Dimer-spin configurations along $\vb*{e}_\parallel$ and $\vb*{e}_\perp$. 
		\textbf{f} Local spin-orbital entanglement entropy $S_{\rm EE}$ and spin and orbital moments, $|\langle\vb*{S}\rangle|$ and $|\langle\vb*{\tau}\rangle|$, as functions of $\kappa/J$.
	}
	\label{fig:incommensurate}
\end{figure*}

\subsection*{Unidirectional incommensurate spiral magnet with spin-orbital entanglement}
The finite-momentum instability of the uniform-dimer configuration at $\kappa_{c_3}$
raises the question of the nature of the ground state beyond this instability. 
To investigate the resulting ground state, we perform a variational calculation, supplemented by a Hessian-based search for lower-energy states (see Methods). 
The resulting state retains the uniform orbital configuration, while the spin configuration develops a long-wavelength coplanar modulation perpendicular to the dimer direction. 

This modulation is visible in the spin structure factor. 
As shown in Fig.~\ref{fig:incommensurate}a, the ordering peak at $\vb*{Q}_0$ of the uniform-dimer configuration splits into a pair of peaks at $\vb*{Q}_0\pm q_\perp\vb*{e}_\perp$. 
Because the allowed momenta are discrete in finite systems, the value of $q_\perp$ that maximizes the structure factor exhibits nonmonotonic finite-size variations, as shown in Fig.~\ref{fig:incommensurate}b. 
Nevertheless, the data approach a finite range with increasing $L$, from which we estimate the modulation wave vector $q_*$ as the midpoint of its upper and lower bounds (see SM for details).
The resulting nonzero $q_*$ demonstrates that the modulation persists in the thermodynamic limit.

Figure~\ref{fig:incommensurate}c compares $q_*$ with $q_{\rm LFW}$, defined as the wave vector that minimizes $\omega_G^2$ of the uniform-dimer configuration within LFW theory. 
Although they differ quantitatively, both increase as $\sqrt{\kappa - \kappa_{c_3}}$ near the transition, supporting the Lifshitz-type commensurate-incommensurate transition anticipated from the LFW instability. 

The real-space spin texture retains the dimer structure of the uniform-dimer configuration.
The two spins within each FM dimer remain parallel and can be represented by a single dimer spin, as shown in Fig.~\ref{fig:incommensurate}d. 
These dimer spins are ferromagnetically aligned along the dimer direction $\vb*{e}_\parallel$, but form a coplanar spiral with wave vector $q_*$ along the perpendicular direction $\vb*{e}_\perp$, as shown in Fig.~\ref{fig:incommensurate}e. 
The spiral can wind in either sense, corresponding to degenerate states with $\pm q_*$, whose selection spontaneously breaks the discrete chirality symmetry. 

The spin modulation also induces local spin-orbital entanglement, consistent with the coupling between spin gradients and spin-orbital fluctuations [Eq.~\eqref{eq:induced_so_fluctuation}].
Figure~\ref{fig:incommensurate}f shows the local spin-orbital entanglement entropy $S_{\rm EE}$ and the magnitudes of the spin and orbital moments. 
In the modulated state, $S_{\rm EE}$ becomes finite and increases with $\kappa/J$, while $|\langle\vb*{S}\rangle|$ and $|\langle\vb*{\tau}\rangle|$ decrease from their separable-state values of $1/2$.
The extraction of these quantities from finite-size calculations is described in SM. 

\subsection*{Microscopic realization of the orbital-field interaction and candidate materials}

We now discuss possible microscopic realizations of the orbital-field model considered here. 
Starting from a multiorbital Hubbard model for $t_{2g}$ orbital systems under a trigonal crystal field, a strong-coupling expansion generally generates the orbital-field interaction together with conventional bilinear orbital exchanges.
Unlike conventional $e_g$ two-orbital systems, where hopping is typically considered within the active orbital doublet, the $e_g^\pi$ doublet also allows hopping to the $a_{1g}$ orbital, making it possible to suppress intra-doublet hopping while retaining substantial interorbital hopping. 
Consequently, the bilinear orbital exchanges can be suppressed and vanish in the absence of intra-doublet hopping, while hopping processes involving the $a_{1g}$ orbital retain the $\kappa$ interaction. 
An analysis based on Slater-Koster parameters shows that such a hopping hierarchy can be realized in edge-sharing octahedra through a suitable balance between direct and ligand-mediated hopping processes.
Details of these analyses are provided in SM.

Deviations from this idealized limit generate additional orbital interactions, but they do not necessarily eliminate the subextensive degeneracy found in the minimal model.
The symmetry-allowed bilinear orbital interactions~\cite{Nakai2026} can be classified into those that preserve the subextensive degeneracy and those that lift it by selecting either the uniform or alternating dimer configuration depending on their sign.
Spin-orbit coupling also deserves consideration because, despite the relatively weak atomic spin-orbit coupling of $3d$ ions, an active orbital degree of freedom allows it to act significantly on the spin-orbital states~\cite{Kanamori1957, Yang2020_VI3, Nakai2026}.
We find that spin-orbit coupling breaks the spin-rotation symmetry but does not lift the subextensive degeneracy. 
Beyond their effects on degeneracy, we also examine whether these additional interactions destabilize the phases of the minimal model. 
Variational calculations show that the phase structure remains robust against perturbations of appreciable strength, with competing phases emerging only when their magnitudes reach approximately $0.3\kappa$ or larger, depending on the interaction (see SM for details). 

Both $d^1$ and $d^2$ electronic configurations provide possible settings for this exchange mechanism, although the resulting balance between Heisenberg and orbital-field interactions depends on the electronic configuration and microscopic parameters.
In particular, for the $d^2$ configuration, the strong-coupling expansion shows that Heisenberg interaction can change sign with increasing Hund's coupling and pass through the pure-orbital-field limit.
Candidate material families include $d^1$ vanadates $A$VO$_3$ with the ilmenite structure~\cite{Yamamoto2022_MgVO3, Solana-Madruga2023_MnVO3, Yamamoto2024_ZnVO3, Yamamoto2026_CuVO3} and $d^2$ vanadium trihalides V$X_3$~\cite{Mastrippolito2023_VCl3, Kao2026_VCl3, Kong2019_VBr3, Kao2026_VBr3, Kong2019_VI3, Hao2021_VI3, Gu2024_VI3}.
Although real compounds generally contain additional interactions and structural distortions and often develop magnetic and orbital order at low temperatures, the above analysis suggests that the frustration mechanism identified here may remain relevant when the orbital-field interaction is sufficiently prominent.
Chemical substitution, pressure, and strain provide possible means of controlling the microscopic hopping processes and the relative strengths of the resulting spin-orbital interactions.
Indeed, recent first-principles studies of VCl$_3$ have demonstrated a strong strain dependence of its exchange interactions~\cite{Camerano2025_VCl3}, illustrating the tunability in this class of materials.

\section*{Discussion}

In summary, we have identified a mechanism of magnetic frustration arising from intrinsic orbital fields.
Because spin correlations on different bonds contribute to the orbital field along different directions in pseudospin space, the spin and orbital configurations cannot be simultaneously optimized.
The resulting local constraints produce a subextensively degenerate ground state manifold and an effective dimensional reduction into independently configurable zigzag chains.
Quantum fluctuations lift this degeneracy and select distinct spin-orbital orders.
At large orbital-field interaction, the spin-orbital fluctuations soften the spin Goldstone mode at finite momentum and eventually destabilize the uniform-dimer configuration.
Beyond this instability, an incommensurate coplanar spiral emerges through a Lifshitz-type transition, accompanied by local spin-orbital entanglement. 

More broadly, intrinsic orbital fields generated by spin correlations can arise naturally in orbitally degenerate systems described by Kugel-Khomskii models, and are therefore not specific to the honeycomb lattice considered here. Despite their potentially important role, such couplings have received relatively little attention.
Through bond-dependent orbital anisotropy, such couplings may generate local constraints in a broader class of spin-orbital systems. Local constraints often underlie emergent gauge fields and fractionalized magnetic excitations~\cite{Castelnovo2008, Henley2010, Yan2024}. Although such phenomena lie beyond the semiclassical approach employed here, they raise the possibility of unconventional quantum phases arising from the spin-orbital constraints identified in this work. Exploring these possibilities in the fully quantum model remains an important direction for future research.

\section*{Methods}

\subsection*{Variational calculation}

We employ a site-factorized variational wave function, $\ket*{\Psi}=\prod_i\ket*{\Omega_i}$, with each local state parameterized by six real variables as an SU(4) coherent state~\cite{Nemoto2000}. 
This ansatz includes local spin-orbital entanglement but neglects intersite entanglement. 
The energy is minimized using the L-BFGS algorithm implemented in Optim.jl~\cite{Mogensen2018}.

We use multiple initial configurations to explore the variational energy landscape, which contains numerous local minima. 
Beyond $\kappa_{c_3}$, where the uniform-dimer configuration becomes a saddle point, we identify negative eigenmodes of the Hessian with respect to local SU(4) variations, perturb the state along both directions of the lowest mode, and reoptimize until no negative Hessian eigenvalues are found. 

\subsection*{Linear flavor-wave theory}

We perform LFW calculations around classical spin-orbital product states.
Since the local Hilbert space is four dimensional, we introduce three flavor-wave bosons, $b_{i,s}$, $b_{i,o}$, and $b_{i,so}$, describing spin, orbital, and spin-orbital fluctuations, respectively, relative to the local reference state~\cite{Joshi1999}.
Expanding the Hamiltonian to quadratic order and applying a bosonic Bogoliubov transformation yields the excitation spectrum and zero-point energy. 

For the dimer configurations, the quadratic Hamiltonian is expressed as  
\begin{equation}\label{eq:LFW_Ham}
\begin{split}
\mathcal{H}^{(2)}
&= \sum_i \left(
\epsilon_s b_{i,s}^{\dagger}b_{i,s}
+\epsilon_o b_{i,o}^{\dagger}b_{i,o}
+\epsilon_{so} b_{i,so}^{\dagger}b_{i,so}
\right)\\
&\quad
+t\sum_{\langle ij\rangle_{\rm FM}}
\left(b_{i,s}^{\dagger}b_{j,s}+\mathrm{H.c.}\right)\\
&\quad
+\Delta\sum_{\langle ij\rangle_{\rm AFM}}
\left(b_{i,s}^{\dagger}b_{j,s}^{\dagger}+\mathrm{H.c.}\right)\\
&\quad
-\frac{\kappa}{2}\sum_{\langle ij\rangle_{\rm AFM}}
\Big[
s_j^\gamma b_{i,s}^{\dagger}b_{j,so}^{\dagger}
+s_i^\gamma b_{i,so}^{\dagger}b_{j,s}^{\dagger}
+\mathrm{H.c.}
\Big],
\end{split}
\end{equation}
where $s_i^\gamma=\sin(\phi_\gamma-\alpha_i)$ and $\alpha_i$ is the orbital pseudospin angle measured from the positive $\tau^x$ axis in the $\tau^x$-$\tau^y$ plane at site $i$. 
The coefficients are given by
\begin{equation}
\begin{split}
&\epsilon_s=\frac{J}{2}+2\kappa, \quad
\epsilon_o=\kappa, \quad
\epsilon_{so}=\frac{J}{2}+\kappa, \\
&t=\frac{J-2\kappa}{2},\quad
\Delta=-\frac{J+\kappa}{2}.
\end{split}
\end{equation}
The dependence on the dimer configuration enters through the arrangement of FM and AFM bonds and the orbital coefficients $s_i^\gamma$.
The derivation of this quadratic Hamiltonian is provided in SM. 

The amplitudes of the spin hopping and pairing terms depend on both $J$ and $\kappa$, reflecting how the $\kappa$ interaction modulates the effective spin exchange through the orbital configuration.
Furthermore, the $\kappa$ interaction also generates hybridization between the spin and spin-orbital fluctuations. 

\section*{Data availability}
The datasets generated and/or analyzed during the current study are not publicly available because they have not been prepared for public distribution but are available from the corresponding author on reasonable request.

\section*{Code availability}
The code used in this study is not publicly available because it has not been prepared for public distribution but is available from the corresponding author on reasonable request.

\begin{acknowledgments}
	H.-Y. K. thanks G. Andolfatto and J. Cen for their contributions during the early stages of this work. 
\end{acknowledgments}

\section*{Funding}

This work is supported by the NSERC Discovery Grant No. 2022-04601. 
H.N. was supported by JSPS Overseas Research Fellowship.
H.-Y. K. acknowledges support from the Canada Research Chairs Program No. CRC-2019-00147.
This research was enabled in part by support provided by Compute Ontario, Calcul Québec, and the Digital Research Alliance of Canada.
The funders had no role in designing or conceiving the project, the calculations and analysis, or writing the manuscript.

\section*{Competing interests}
The authors declare no competing financial or non-financial interests.

\section*{Author contributions}
H.-Y. K. conceived and supervised the project. H.N. performed the variational calculations and linear flavor-wave theory and analyzed the results. Both authors contributed to writing the manuscript.

\bibliography{refs}

@book{Lacroix2011,
	editor = {Lacroix, Claudine and Mendels, Philippe and Mila, Frédéric},
	title = {Introduction to {Frustrated} {Magnetism}: {Materials}, {Experiments}, {Theory}},
	series = {Springer {Series} in {Solid}-{State} {Sciences}},
	volume = {164},
	publisher = {Springer},
	address   = {Berlin, Heidelberg},
	url = {https://link.springer.com/10.1007/978-3-642-10589-0},
	doi = {10.1007/978-3-642-10589-0},
	year = {2011},
}

@article{Castelnovo2008,
	title = {{Magnetic monopoles in spin ice}},
	volume = {451},
	copyright = {http://www.springer.com/tdm},
	issn = {0028-0836, 1476-4687},
	url = {https://www.nature.com/articles/nature06433},
	doi = {10.1038/nature06433},
	number = {7174},
	journal = {Nature},
	author = {Castelnovo, C. and Moessner, R. and Sondhi, S. L.},
	month = jan,
	year = {2008},
	pages = {42},
}

@article{Henley2010,
	title = {{The ``Coulomb Phase'' in Frustrated Systems}},
	volume = {1},
	issn = {1947-5454, 1947-5462},
	url = {https://www.annualreviews.org/doi/10.1146/annurev-conmatphys-070909-104138},
	doi = {10.1146/annurev-conmatphys-070909-104138},
	number = {1},
	journal = {Annu. Rev. Condens. Matter Phys.},
	author = {Henley, Christopher L.},
	month = aug,
	year = {2010},
	pages = {179--210},
}

@article{Yan2024,
	title = {{Classification of classical spin liquids: Detailed formalism and suite of examples}},
	author = {Yan, Han and Benton, Owen and Nevidomskyy, Andriy H. and Moessner, Roderich},
	journal = {Phys. Rev. B},
	volume = {109},
	issue = {17},
	pages = {174421},
	numpages = {35},
	year = {2024},
	month = {May},
	publisher = {American Physical Society},
	doi = {10.1103/PhysRevB.109.174421},
	url = {https://link.aps.org/doi/10.1103/PhysRevB.109.174421}
}

@article{Kugel1982,
	title = {{The Jahn-Teller effect and magnetism: transition metal compounds}},
	volume = {136},
	issn = {0042-1294, 1996-6652},
	url = {http://ufn.ru/ru/articles/1982/4/c/},
	doi = {10.3367/UFNr.0136.198204c.0621},
	number = {4},
	journal = {Usp. Fiz. Nauk},
	author = {Kugel, Kliment I. and Khomskii, Daniel I.},
	year = {1982},
	pages = {621},
}

@article{Tokura2000,
	title = {{Orbital Physics in Transition-Metal Oxides}},
	volume = {288},
	issn = {0036-8075, 1095-9203},
	url = {https://www.science.org/doi/10.1126/science.288.5465.462},
	doi = {10.1126/science.288.5465.462},
	number = {5465},
	journal = {Science},
	author = {Tokura, Y. and Nagaosa, N.},
	month = apr,
	year = {2000},
	pages = {462},
}

@article{Khaliullin2005,
	title = {{Orbital Order and Fluctuations in Mott Insulators}},
	volume = {160},
	issn = {0375-9687},
	url = {https://academic.oup.com/ptps/article-lookup/doi/10.1143/PTPS.160.155},
	doi = {10.1143/PTPS.160.155},
	journal = {Prog. Theor. Phys. Suppl.},
	author = {Khaliullin, Giniyat},
	year = {2005},
	pages = {155},
}

@article{Streltsov2017,
	title = {{Orbital physics in transition metal compounds: new trends}},
	volume = {60},
	issn = {1063-7869, 1468-4780},
	url = {https://ufn.ru/en/articles/2017/11/d/},
	doi = {10.3367/UFNe.2017.08.038196},
	number = {11},
	journal = {Phys.-Usp.},
	author = {Streltsov, S V and Khomskii, D I},
	month = nov,
	year = {2017},
	pages = {1121},
}

@article{Yamamoto2022_MgVO3,
	title = {{Cation Dimerization in a 3d$^1$ Honeycomb Lattice System}},
	volume = {144},
	copyright = {https://doi.org/10.15223/policy-029},
	issn = {0002-7863, 1520-5126},
	url = {https://pubs.acs.org/doi/10.1021/jacs.1c10977},
	doi = {10.1021/jacs.1c10977},
	number = {3},
	journal = {J. Am. Chem. Soc.},
	author = {Yamamoto, Hajime and Kamiyama, Sachiko and Yamada, Ikuya and Kimura, Hiroyuki},
	month = jan,
	year = {2022},
	pages = {1082--1086},
}

@article{Solana-Madruga2023_MnVO3,
	title = {{High-pressure ilmenite-type MnVO$_3$ : crystal and spin structures in the itinerant-localized regimes}},
	volume = {11},
	issn = {2050-7526, 2050-7534},
	url = {https://pubs.rsc.org/tc/article/11/27/9238-9244/834547},
	doi = {10.1039/D3TC01541F},
	number = {27},
	journal = {J. Mater. Chem. C},
	author = {Solana-Madruga, Elena and Mentré, Olivier and Arévalo-López, Eugenia P. and Khalyavin, Dmitry and Fauth, Francois and Missiul, Alexandr and Arévalo-López, Angel M.},
	year = {2023},
	pages = {9238--9244},
}

@article{Yamamoto2024_ZnVO3,
	title = {{ZnVO$_3$ : an ilmenite-type vanadium oxide hosting robust V-V dimers}},
	volume = {53},
	issn = {1477-9226, 1477-9234},
	url = {https://pubs.rsc.org/dt/article/53/39/16195-16201/847366},
	doi = {10.1039/D4DT02239D},
	number = {39},
	journal = {Dalton Trans.},
	author = {Yamamoto, Hajime and Nishikubo, Takumi and Kobayashi, Shintaro and Takahashi, Kazuki and Azuma, Masaki and Kawaguchi, Shogo and Abukawa, Tadashi},
	year = {2024},
	pages = {16195--16201},
}

@article{Yamamoto2026_CuVO3,
	title = {{Spin and Orbital States in Triclinic Ilmenite-Type CuVO$_3$}},
	volume = {11},
	copyright = {https://creativecommons.org/licenses/by/4.0/},
	issn = {2470-1343, 2470-1343},
	url = {https://pubs.acs.org/doi/10.1021/acsomega.5c13303},
	doi = {10.1021/acsomega.5c13303},
	number = {14},
	journal = {ACS Omega},
	author = {Yamamoto, Hajime and Ochi, Keigo and Aoyama, Takuya and Ishii, Kenji and Matsumura, Daiju and Tsuji, Takuya and Ohgushi, Kenya and Abukawa, Tadashi},
	month = apr,
	year = {2026},
	pages = {22093--22098},
}

@article{Mastrippolito2023_VCl3,
	title = {{Polaronic and Mott insulating phase of layered magnetic vanadium trihalide ${\mathrm{VCl}}_{3}$}},
	author = {Mastrippolito, Dario and Camerano, Luigi and {\'S}wi{\k a}tek, Hanna and {\v S}m{\'\i}d, B{\v r}etislav and Klimczuk, Tomasz and Ottaviano, Luca and Profeta, Gianni},
	journal = {Phys. Rev. B},
	volume = {108},
	issue = {4},
	pages = {045126},
	numpages = {7},
	year = {2023},
	month = {Jul},
	publisher = {American Physical Society},
	doi = {10.1103/PhysRevB.108.045126},
	url = {https://link.aps.org/doi/10.1103/PhysRevB.108.045126}
}

@article{Camerano2025_VCl3,
	title = {{Multicomponent Magneto-Orbital Order and Magneto-Orbitons in Monolayer ${\mathrm{VCl}}_{3}$}},
	volume = {25},
	copyright = {https://creativecommons.org/licenses/by/4.0/},
	issn = {1530-6984, 1530-6992},
	url = {https://pubs.acs.org/doi/10.1021/acs.nanolett.4c06400},
	doi = {10.1021/acs.nanolett.4c06400},
	number = {12},
	journal = {Nano Lett.},
	author = {Camerano, Luigi and Fumega, Adolfo O. and Profeta, Gianni and Lado, Jose L.},
	month = mar,
	year = {2025},
	pages = {4825--4831},
}

@article{Kao2026_VCl3,
	title = {{Magnetic order in the van der Waals magnet ${\mathrm{VCl}}_{3}$}},
	author = {Kao, Zeyu and Hao, Yiqing and Gu, Yimeng and Chen, Lixing and Zhang, Enkang and Zhang, Hao and Orlandi, Fabio and Manuel, Pascal and Wang, Junfeng and Dong, Chao and Xi, Chuanying and Chen, Zefeng and Xu, Changsong and Zhao, Jun},
	journal = {Phys. Rev. B},
	volume = {113},
	issue = {9},
	pages = {094428},
	numpages = {9},
	year = {2026},
	month = {Mar},
	publisher = {American Physical Society},
	doi = {10.1103/fj11-mwxv},
	url = {https://link.aps.org/doi/10.1103/fj11-mwxv}
}

@article{Kong2019_VBr3,
	title = {{Crystal structure and magnetic properties of the layered van der Waals compound $\mathrm{VB}{\mathrm{r}}_{3}$}},
	author = {Kong, Tai and Guo, Shu and Ni, Danrui and Cava, Robert J.},
	journal = {Phys. Rev. Mater.},
	volume = {3},
	issue = {8},
	pages = {084419},
	numpages = {5},
	year = {2019},
	month = {Aug},
	publisher = {American Physical Society},
	doi = {10.1103/PhysRevMaterials.3.084419},
	url = {https://link.aps.org/doi/10.1103/PhysRevMaterials.3.084419}
}

@article{Kao2026_VBr3,
	title = {{Kitaev interactions in the van der Waals antiferromagnet $\mathrm{VB}{\mathrm{r}}_{3}$}},
	volume = {71},
	issn = {20959273},
	url = {https://linkinghub.elsevier.com/retrieve/pii/S2095927326000654},
	doi = {10.1016/j.scib.2026.01.040},
	number = {6},
	journal = {Science Bulletin},
	author = {Kao, Zeyu and Gu, Yimeng and Gu, Yiqing and Zhang, Hao and Zheng, Shiyi and Murai, Naoki and Ohira-Kawamura, Seiko and Zhao, Jun},
	month = mar,
	year = {2026},
	pages = {1335--1340},
}

@article{Kong2019_VI3,
	title = {{VI$_3$—a New Layered Ferromagnetic Semiconductor}},
	volume = {31},
	issn = {0935-9648, 1521-4095},
	url = {https://advanced.onlinelibrary.wiley.com/doi/10.1002/adma.201808074},
	doi = {10.1002/adma.201808074},
	number = {17},
	journal = {Adv. Mater.},
	author = {Kong, Tai and Stolze, Karoline and Timmons, Erik I. and Tao, Jing and Ni, Danrui and Guo, Shu and Yang, Zoë and Prozorov, Ruslan and Cava, Robert J.},
	month = apr,
	year = {2019},
	pages = {1808074},
}

@article{Yang2020_VI3,
	title = {{${\mathrm{VI}}_{3}$: A two-dimensional Ising ferromagnet}},
	author = {Yang, Ke and Fan, Fengren and Wang, Hongbo and Khomskii, D. I. and Wu, Hua},
	journal = {Phys. Rev. B},
	volume = {101},
	issue = {10},
	pages = {100402(R)},
	numpages = {5},
	year = {2020},
	month = {Mar},
	publisher = {American Physical Society},
	doi = {10.1103/PhysRevB.101.100402},
	url = {https://link.aps.org/doi/10.1103/PhysRevB.101.100402}
}

@article{Hao2021_VI3,
	title = {{Magnetic Order and Its Interplay with Structure Phase Transition in van der Waals Ferromagnet VI$_3$}},
	volume = {38},
	issn = {0256-307X, 1741-3540},
	url = {https://iopscience.iop.org/article/10.1088/0256-307X/38/9/096101},
	doi = {10.1088/0256-307X/38/9/096101},
	number = {9},
	journal = {Chinese Phys. Lett.},
	author = {Hao, Yiqing and Gu, Yiqing and Gu, Yimeng and Feng, Erxi and Cao, Huibo and Chi, Songxue and Wu, Hua and Zhao, Jun},
	month = oct,
	year = {2021},
	pages = {096101},
}

@article{Gu2024_VI3,
	title = {{Signatures of Kitaev Interactions in the van der Waals Ferromagnet ${\mathrm{VI}}_{3}$}},
	author = {Gu, Yiqing and Gu, Yimeng and Liu, Feiyang and Ohira-Kawamura, Seiko and Murai, Naoki and Zhao, Jun},
	journal = {Phys. Rev. Lett.},
	volume = {132},
	issue = {24},
	pages = {246702},
	numpages = {7},
	year = {2024},
	month = {Jun},
	publisher = {American Physical Society},
	doi = {10.1103/PhysRevLett.132.246702},
	url = {https://link.aps.org/doi/10.1103/PhysRevLett.132.246702}
}

@article{Feiner1997,
	title = {{Quantum Melting of Magnetic Order due to Orbital Fluctuations}},
	author = {Feiner, Louis Felix and Ole\ifmmode \acute{s}\else \'{s}\fi{}, Andrzej M. and Zaanen, Jan},
	journal = {Phys. Rev. Lett.},
	volume = {78},
	issue = {14},
	pages = {2799},
	numpages = {0},
	year = {1997},
	month = {Apr},
	publisher = {American Physical Society},
	doi = {10.1103/PhysRevLett.78.2799},
	url = {https://link.aps.org/doi/10.1103/PhysRevLett.78.2799}
}

@article{Khaliullin2000,
	title = {{Orbital Liquid in Three-Dimensional Mott Insulator: ${\mathrm{LaTiO}}_{3}$}},
	author = {Khaliullin, G. and Maekawa, S.},
	journal = {Phys. Rev. Lett.},
	volume = {85},
	issue = {18},
	pages = {3950},
	numpages = {0},
	year = {2000},
	month = {Oct},
	publisher = {American Physical Society},
	doi = {10.1103/PhysRevLett.85.3950},
	url = {https://link.aps.org/doi/10.1103/PhysRevLett.85.3950}
}

@article{Vernay2004,
	title = {{Orbital degeneracy as a source of frustration in ${\mathrm{LiNiO}}_{2}$}},
	author = {Vernay, F. and Penc, K. and Fazekas, P. and Mila, F.},
	journal = {Phys. Rev. B},
	volume = {70},
	issue = {1},
	pages = {014428},
	numpages = {17},
	year = {2004},
	month = {Jul},
	publisher = {American Physical Society},
	doi = {10.1103/PhysRevB.70.014428},
	url = {https://link.aps.org/doi/10.1103/PhysRevB.70.014428}
}

@article{Normand2008,
	title = {{Frustration and entanglement in the ${t}_{2g}$ spin-orbital model on a triangular lattice: Valence-bond and generalized liquid states}},
	author = {Normand, Bruce and Ole\ifmmode \acute{s}\else \'{s}\fi{}, Andrzej M.},
	journal = {Phys. Rev. B},
	volume = {78},
	issue = {9},
	pages = {094427},
	numpages = {39},
	year = {2008},
	month = {Sep},
	publisher = {American Physical Society},
	doi = {10.1103/PhysRevB.78.094427},
	url = {https://link.aps.org/doi/10.1103/PhysRevB.78.094427}
}

@article{Savary2021,
	title = {{Quantum loop states in spin-orbital models on the honeycomb lattice}},
	volume = {12},
	issn = {2041-1723},
	url = {https://www.nature.com/articles/s41467-021-23033-y},
	doi = {10.1038/s41467-021-23033-y},
	number = {1},
	journal = {Nat. Commun.},
	author = {Savary, Lucile},
	month = may,
	year = {2021},
	pages = {3004},
}

@article{Churchill2025,
	title = {{Microscopic roadmap to a Kitaev-Yao-Lee spin-orbital liquid}},
	volume = {10},
	issn = {2397-4648},
	url = {https://www.nature.com/articles/s41535-025-00744-9},
	doi = {10.1038/s41535-025-00744-9},
	number = {1},
	journal = {npj Quantum Mater.},
	author = {Churchill, Derek and Zhang, Emily Z. and Kee, Hae-Young},
	month = mar,
	year = {2025},
	pages = {26},
}

@article{Kugel1973,
	title = {{Crystal structure and magnetic properties of substances with orbital degeneracy}},
	author = {Kugel, K.~I. and Khomskii, D.~I.},
	journal = {Sov. Phys. JETP},
	year = {1973},
	month = {Oct},
	volume = {37},
	pages = {725}
}

@article{Kugel2015,
	title = {{Spin-orbital interaction for face-sharing octahedra: Realization of a highly symmetric SU(4) model}},
	author = {Kugel, K. I. and Khomskii, D. I. and Sboychakov, A. O. and Streltsov, S. V.},
	journal = {Phys. Rev. B},
	volume = {91},
	issue = {15},
	pages = {155125},
	numpages = {11},
	year = {2015},
	month = {Apr},
	publisher = {American Physical Society},
	doi = {10.1103/PhysRevB.91.155125},
	url = {https://link.aps.org/doi/10.1103/PhysRevB.91.155125}
}

@article{Chern2010,
	title = {{Quantum $120\ifmmode^\circ\else\textdegree\fi{}$ model on pyrochlore lattice: Orbital ordering in ${\text{MnV}}_{2}{\text{O}}_{4}$}},
	author = {Chern, Gia-Wei and Perkins, Natalia and Hao, Zhihao},
	journal = {Phys. Rev. B},
	volume = {81},
	issue = {12},
	pages = {125127},
	numpages = {10},
	year = {2010},
	month = {Mar},
	publisher = {American Physical Society},
	doi = {10.1103/PhysRevB.81.125127},
	url = {https://link.aps.org/doi/10.1103/PhysRevB.81.125127}
}

@misc{Nakai2026,
	title={{Spin-orbital exchange as a route to intertwined dipole-quadrupole orbital order in MnV$_2$O$_4$ under strong trigonal crystal field}}, 
	author={Hiroki Nakai and Yusuke Nomura},
	year={2026},
	eprint={2605.03028},
	archivePrefix={arXiv},
	primaryClass={cond-mat.str-el},
	url={https://arxiv.org/abs/2605.03028}, 
}

@article{Jackeli2008,
	title = {{Classical Dimers and Dimerized Superstructure in an Orbitally Degenerate Honeycomb Antiferromagnet}},
	author = {Jackeli, G. and Khomskii, D. I.},
	journal = {Phys. Rev. Lett.},
	volume = {100},
	issue = {14},
	pages = {147203},
	numpages = {4},
	year = {2008},
	month = {Apr},
	publisher = {American Physical Society},
	doi = {10.1103/PhysRevLett.100.147203},
	url = {https://link.aps.org/doi/10.1103/PhysRevLett.100.147203}
}

@article{Brzezicki2012,
	title = {{Noncollinear Magnetic Order Stabilized by Entangled Spin-Orbital Fluctuations}},
	author = {Brzezicki, Wojciech and Dziarmaga, Jacek and Ole{\ifmmode \acute{s}\else \'{s}\fi{}}, Andrzej M.},
	journal = {Phys. Rev. Lett.},
	volume = {109},
	issue = {23},
	pages = {237201},
	numpages = {5},
	year = {2012},
	month = {Dec},
	publisher = {American Physical Society},
	doi = {10.1103/PhysRevLett.109.237201},
	url = {https://link.aps.org/doi/10.1103/PhysRevLett.109.237201}
}

@article{Khomskii2003,
	title = {{Orbital ordering and frustrations}},
	volume = {36},
	issn = {0305-4470, 1361-6447},
	url = {https://iopscience.iop.org/article/10.1088/0305-4470/36/35/307},
	doi = {10.1088/0305-4470/36/35/307},
	number = {35},
	journal = {J. Phys. A: Math. Gen.},
	author = {Khomskii, D I and Mostovoy, M V},
	month = sep,
	year = {2003},
	pages = {9197--9207},
}

@article{Mishra2004,
	title = {{Directional Ordering of Fluctuations in a Two-Dimensional Compass Model}},
	author = {Mishra, Anup and Ma, Michael and Zhang, Fu-Chun and Guertler, Siegfried and Tang, Lei-Han and Wan, Shaolong},
	journal = {Phys. Rev. Lett.},
	volume = {93},
	issue = {20},
	pages = {207201},
	numpages = {4},
	year = {2004},
	month = {Nov},
	publisher = {American Physical Society},
	doi = {10.1103/PhysRevLett.93.207201},
	url = {https://link.aps.org/doi/10.1103/PhysRevLett.93.207201}
}

@article{Biskup2005,
	title = {{Orbital Ordering in Transition-Metal Compounds: I. The 120-Degree Model}},
	volume = {255},
	copyright = {http://www.springer.com/tdm},
	issn = {0010-3616, 1432-0916},
	url = {http://link.springer.com/10.1007/s00220-004-1272-7},
	doi = {10.1007/s00220-004-1272-7},
	number = {2},
	journal = {Commun. Math. Phys.},
	author = {Biskup, M. and Chayes, L. and Nussinov, Z.},
	month = apr,
	year = {2005},
	pages = {253--292},
}

@article{Nussinov2015,
	title = {{Compass models: Theory and physical motivations}},
	author = {Nussinov, Zohar and van den Brink, Jeroen},
	journal = {Rev. Mod. Phys.},
	volume = {87},
	issue = {1},
	pages = {1--59},
	numpages = {59},
	year = {2015},
	month = {Jan},
	publisher = {American Physical Society},
	doi = {10.1103/RevModPhys.87.1},
	url = {https://link.aps.org/doi/10.1103/RevModPhys.87.1}
}

@article{Wu2008,
	title = {{Orbital Ordering and Frustration of $p$-Band Mott Insulators}},
	author = {Wu, Congjun},
	journal = {Phys. Rev. Lett.},
	volume = {100},
	issue = {20},
	pages = {200406},
	numpages = {4},
	year = {2008},
	month = {May},
	publisher = {American Physical Society},
	doi = {10.1103/PhysRevLett.100.200406},
	url = {https://link.aps.org/doi/10.1103/PhysRevLett.100.200406}
}

@article{Nasu2008,
	title = {{Doubly degenerate orbital system in honeycomb lattice: Implication of orbital state in layered iron oxide}},
	author = {Nasu, J. and Nagano, A. and Naka, M. and Ishihara, S.},
	journal = {Phys. Rev. B},
	volume = {78},
	issue = {2},
	pages = {024416},
	numpages = {12},
	year = {2008},
	month = {Jul},
	publisher = {American Physical Society},
	doi = {10.1103/PhysRevB.78.024416},
	url = {https://link.aps.org/doi/10.1103/PhysRevB.78.024416}
}

@article{Hornreich1975,
	title = {{Critical Behavior at the Onset of $\stackrel{\ensuremath{\rightarrow}}{\mathrm{k}}$-Space Instability on the $\ensuremath{\lambda}$ Line}},
	author = {Hornreich, R. M. and Luban, Marshall and Shtrikman, S.},
	journal = {Phys. Rev. Lett.},
	volume = {35},
	issue = {25},
	pages = {1678--1681},
	numpages = {0},
	year = {1975},
	month = {Dec},
	publisher = {American Physical Society},
	doi = {10.1103/PhysRevLett.35.1678},
	url = {https://link.aps.org/doi/10.1103/PhysRevLett.35.1678}
}

@article{Redner1977,
	title = {{Helical order and its onset at the Lifshitz point}},
	author = {Redner, S. and Stanley, H. E.},
	journal = {Phys. Rev. B},
	volume = {16},
	issue = {11},
	pages = {4901--4906},
	numpages = {0},
	year = {1977},
	month = {Dec},
	publisher = {American Physical Society},
	doi = {10.1103/PhysRevB.16.4901},
	url = {https://link.aps.org/doi/10.1103/PhysRevB.16.4901}
}

@article{Kitaev2006,
	title = {{Anyons in an exactly solved model and beyond}},
	volume = {321},
	issn = {00034916},
	url = {https://linkinghub.elsevier.com/retrieve/pii/S0003491605002381},
	doi = {10.1016/j.aop.2005.10.005},
	number = {1},
	journal = {Ann. Phys.},
	author = {Kitaev, Alexei},
	month = jan,
	year = {2006},
	pages = {2},
}

@article{Jackeli2009,
	title = {{Mott Insulators in the Strong Spin-Orbit Coupling Limit: From Heisenberg to a Quantum Compass and Kitaev Models}},
	author = {Jackeli, G. and Khaliullin, G.},
	journal = {Phys. Rev. Lett.},
	volume = {102},
	issue = {1},
	pages = {017205},
	numpages = {4},
	year = {2009},
	month = {Jan},
	publisher = {American Physical Society},
	doi = {10.1103/PhysRevLett.102.017205},
	url = {https://link.aps.org/doi/10.1103/PhysRevLett.102.017205}
}

@article{Chaloupka2010,
	title = {{Kitaev-Heisenberg Model on a Honeycomb Lattice: Possible Exotic Phases in Iridium Oxides ${A}_{2}{\mathrm{IrO}}_{3}$}},
	author = {Chaloupka, J and Jackeli, George and Khaliullin, Giniyat},
	journal = {Phys. Rev. Lett.},
	volume = {105},
	issue = {2},
	pages = {027204},
	numpages = {4},
	year = {2010},
	month = {Jul},
	publisher = {American Physical Society},
	doi = {10.1103/PhysRevLett.105.027204},
	url = {https://link.aps.org/doi/10.1103/PhysRevLett.105.027204}
}

@article{Rau2014,
	title = {{Generic Spin Model for the Honeycomb Iridates beyond the Kitaev Limit}},
	author = {Rau, Jeffrey G. and Lee, Eric Kin-Ho and Kee, Hae-Young},
	journal = {Phys. Rev. Lett.},
	volume = {112},
	issue = {7},
	pages = {077204},
	numpages = {5},
	year = {2014},
	month = {Feb},
	publisher = {American Physical Society},
	doi = {10.1103/PhysRevLett.112.077204},
	url = {https://link.aps.org/doi/10.1103/PhysRevLett.112.077204}
}

@article{Wannier1950,
	title = {{Antiferromagnetism. The Triangular Ising Net}},
	author = {Wannier, G. H.},
	journal = {Phys. Rev.},
	volume = {79},
	issue = {2},
	pages = {357--364},
	numpages = {0},
	year = {1950},
	month = {Jul},
	publisher = {American Physical Society},
	doi = {10.1103/PhysRev.79.357},
	url = {https://link.aps.org/doi/10.1103/PhysRev.79.357}
}

@article{Stephenson1964,
	title = {{Ising-Model Spin Correlations on the Triangular Lattice}},
	volume = {5},
	issn = {0022-2488, 1089-7658},
	url = {https://pubs.aip.org/jmp/article/5/8/1009/231386/Ising-Model-Spin-Correlations-on-the-Triangular},
	doi = {10.1063/1.1704202},
	number = {8},
	journal = {J. Math. Phys.},
	author = {Stephenson, John},
	month = aug,
	year = {1964},
	pages = {1009--1024},
}

@article{Moessner2001,
	title = {{Ising models of quantum frustration}},
	author = {Moessner, R. and Sondhi, S. L.},
	journal = {Phys. Rev. B},
	volume = {63},
	issue = {22},
	pages = {224401},
	numpages = {19},
	year = {2001},
	month = {May},
	publisher = {American Physical Society},
	doi = {10.1103/PhysRevB.63.224401},
	url = {https://link.aps.org/doi/10.1103/PhysRevB.63.224401}
}

@article{Slater1954,
	title = {{Simplified LCAO Method for the Periodic Potential Problem}},
	author = {Slater, J. C. and Koster, G. F.},
	journal = {Phys. Rev.},
	volume = {94},
	issue = {6},
	pages = {1498--1524},
	numpages = {0},
	year = {1954},
	month = {Jun},
	publisher = {American Physical Society},
	doi = {10.1103/PhysRev.94.1498},
	url = {https://link.aps.org/doi/10.1103/PhysRev.94.1498}
}

@article{Nemoto2000,
	title = {{Generalized coherent states for \textit{SU}(\textit{n}) systems}},
	volume = {33},
	issn = {0305-4470, 1361-6447},
	url = {https://iopscience.iop.org/article/10.1088/0305-4470/33/17/307},
	doi = {10.1088/0305-4470/33/17/307},
	number = {17},
	journal = {J. Phys. A: Math. Gen.},
	author = {Nemoto, Kae},
	month = may,
	year = {2000},
	pages = {3493},
}

@article{Mogensen2018,
	author  = {Mogensen, Patrick Kofod and Riseth, Asbj{\o}rn Nilsen},
	title   = {{Optim: A mathematical optimization package for Julia}},
	journal = {J. Open Source Softw.},
	year    = {2018},
	volume  = {3},
	number  = {24},
	pages   = {615},
	doi     = {10.21105/joss.00615}
}

@article{Joshi1999,
	title = {{Elementary excitations in magnetically ordered systems with orbital degeneracy}},
	author = {Joshi, A. and Ma, M. and Mila, F. and Shi, D. N. and Zhang, F. C.},
	journal = {Phys. Rev. B},
	volume = {60},
	issue = {9},
	pages = {6584--6587},
	numpages = {0},
	year = {1999},
	month = {Sep},
	publisher = {American Physical Society},
	doi = {10.1103/PhysRevB.60.6584},
	url = {https://link.aps.org/doi/10.1103/PhysRevB.60.6584}
}

@article{Kanamori1957,
	title = {{Theory of the Magnetic Properties of Ferrous and Cobaltous Oxides, I}},
	volume = {17},
	issn = {0033-068X},
	url = {https://academic.oup.com/ptp/article-lookup/doi/10.1143/PTP.17.177},
	doi = {10.1143/PTP.17.177},
	number = {2},
	journal = {Prog. Theor. Phys.},
	author = {Kanamori, Junjiro},
	month = feb,
	year = {1957},
	pages = {177},
}

@article{Kanamori1963,
	title = {{Electron Correlation and Ferromagnetism of Transition Metals}},
	volume = {30},
	issn = {0033-068X},
	url = {https://academic.oup.com/ptp/article-lookup/doi/10.1143/PTP.30.275},
	doi = {10.1143/PTP.30.275},
	number = {3},
	journal = {Prog. Theor. Phys.},
	author = {Kanamori, Junjiro},
	month = sep,
	year = {1963},
	pages = {275},
}

@article{Georges2013,
	title = {{Strong Correlations from Hund's Coupling}},
	volume = {4},
	issn = {1947-5454, 1947-5462},
	url = {https://www.annualreviews.org/doi/10.1146/annurev-conmatphys-020911-125045},
	doi = {10.1146/annurev-conmatphys-020911-125045},
	number = {1},
	journal = {Annu. Rev. Condens. Matter Phys.},
	author = {Georges, Antoine and Medici, Luca De' and Mravlje, Jernej},
	month = apr,
	year = {2013},
	pages = {137},
}
	
\end{document}


\renewcommand{\thefigure}{S\arabic{figure}}
\renewcommand{\thetable}{S\arabic{table}}
\renewcommand{\thesection}{S\arabic{section}}
\renewcommand{\theequation}{S\arabic{equation}}

\begin{center}
	{\large \textmd{Supplementary Materials:} \\[0.3em]
		{\bfseries A cascade of orbital-field-driven phases with a Lifshitz-type transition to unidirectional spiral order}}
\end{center}

\tableofcontents

\section{Dimer configurations and degeneracy}

We derive the constraints on the dimer coverings and count the resulting ground-state (GS) degeneracy. 
As introduced in the main text, each dimer represents an FM spin pair, and spins on neighboring dimers must be antiparallel across unoccupied bonds (the interdimer AFM condition). 
We consider an $L_1\times L_2$ system containing $2L_1L_2$ sites under periodic boundary conditions.
For simplicity, we focus on $L_1=L_2=L$ with even $L$.

\begin{figure*}[t]
	\centering
	\includegraphics[width=16cm]{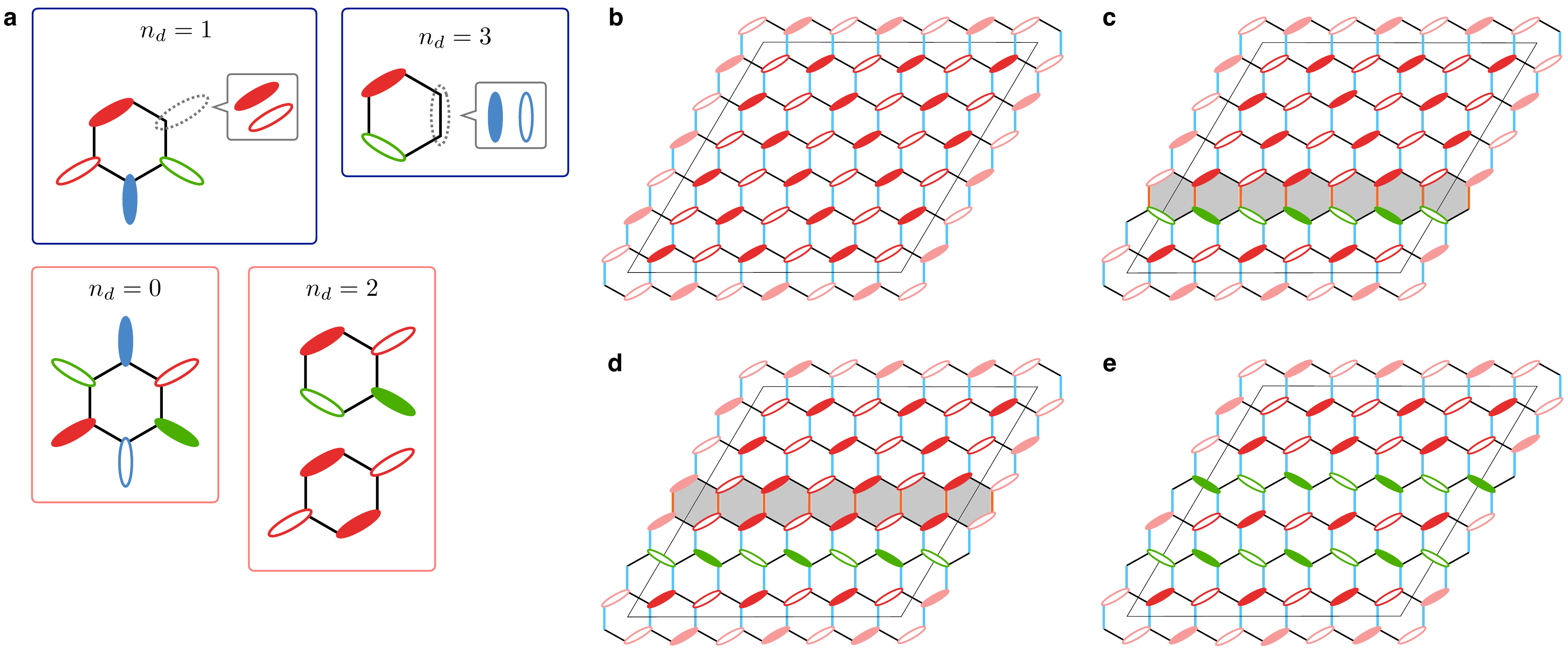}
	\caption{
		\textbf{Dimer configurations and ground-state constraints.}
		\textbf{a} Representative dimer configurations on a hexagonal plaquette with $n_d=0,1,2,3$, where $n_d$ denotes the number of dimers occupying the plaquette edges. Configurations with $n_d=0$ and $2$ satisfy the interdimer AFM condition locally, whereas those with $n_d=1$ and $3$ violate it.
		\textbf{b} Uniform-dimer configuration in which all $xy$-chains are covered by $x$-bond dimers, shown for $L=6$.
		\textbf{c} Configuration obtained by changing the dimer covering on a single $xy$-chain from $x$-bonds to $y$-bonds.
		\textbf{d} Configuration obtained by reversing all spins on a neighboring $x$-dimer chain in \textbf{c}.
		\textbf{e} GS configuration obtained by changing the dimer covering on another $xy$-chain from $x$-bonds to $y$-bonds.
		In \textbf{b}-\textbf{e}, the $z$-bonds connecting neighboring chains are highlighted in blue (AFM) or red (FM), with shaded regions highlighting pairs of chains connected by FM bonds.
	}
	\label{fig:supp_dimer}
\end{figure*}

We first examine the allowed dimer configurations on a hexagonal plaquette.
Let $n_d$ denote the number of dimers occupying its edges ($0\leq n_d\leq 3$).
As shown in Fig.~\ref{fig:supp_dimer}a, the interdimer AFM condition excludes $n_d=1$ and $3$, whereas $n_d=0$ and $2$ are locally allowed.
For $n_d=0$, all six vertices belong to dimers extending outside the plaquette.
Extending this arrangement while satisfying the interdimer AFM condition uniquely determines the dimer covering throughout the lattice, which is incompatible with periodic boundary conditions.
Thus, every hexagon must have $n_d=2$, leading to the three families of parallel zigzag-chain configurations described in the main text.

We now examine the constraints on the dimer patterns of different chains.
Without loss of generality, consider the family of $xy$-chains, each of which can be covered by either $x$-bond or $y$-bond dimers.
For even $L$, the uniform-dimer configuration, in which all chains are covered by $x$-bond dimers, satisfies the GS conditions, as shown in Fig.~\ref{fig:supp_dimer}b for $L=6$.

Starting from this configuration, we change the dimer covering on a single $xy$-chain from $x$-bonds to $y$-bonds, as shown in Fig.~\ref{fig:supp_dimer}c.
The modified chain is connected to its two neighboring chains by $z$-bonds, but the interdimer AFM condition cannot be satisfied on both sets of connecting bonds simultaneously: one set becomes FM while the other remains AFM.
Flipping all spins on one of the neighboring $x$-dimer chains exchanges the FM and AFM connections without eliminating the violation, as shown in Fig.~\ref{fig:supp_dimer}d.
The GS condition can be restored by changing the dimer covering on another $xy$-chain from $x$-bonds to $y$-bonds, as shown in Fig.~\ref{fig:supp_dimer}e.
Repeating this argument shows that periodic boundary conditions require even numbers of both $x$-dimer and $y$-dimer chains.

For a fixed chain direction, this parity constraint leaves $2^{L-1}$ allowed dimer coverings.
There are three equivalent choices of chain direction, but the uniform configurations are counted twice because each belongs to two families.
Removing this overcounting gives $3(2^{L-1}-1)$ distinct dimer coverings.
Each covering admits two spin configurations related by global spin reversal, yielding
\begin{equation}
W_{\rm GS}(L)=6(2^{L-1}-1).
\end{equation}
Therefore, $W_{\rm GS}(L)\sim 2^L$, demonstrating that the GS degeneracy is subextensive.

\section{Linear flavor-wave analysis}

In this section, we derive the quadratic flavor-wave Hamiltonian for the dimer configurations, defined by the dimer coverings satisfying the interdimer AFM condition discussed in the preceding section, and present its application to the uniform-dimer configuration.
We then analyze the Goldstone-mode instability and establish its connection to the phenomenological effective Lagrangian introduced in the main text.

\subsection{Derivation of the quadratic Hamiltonian}

For the dimer configurations, spins are taken to point parallel or antiparallel to the global $z$-axis, while the orbital pseudospins lie in the $\tau^x$-$\tau^y$ plane.
At each site, we introduce local spin and orbital frames aligned with the corresponding classical moments.

For the spin sector, the local $z_{\rm s}$-axis is chosen along the spin direction, while the local $y_{\rm s}$-axis is taken to be common to all sites.
The local frames of antiparallel spins are then related by a $\pi$-rotation about the $y_{\rm s}$-axis.
The spin exchange on a nearest-neighbor bond $\langle ij\rangle$ is thus expressed as
\begin{equation}
\vb*{S}_i\cdot\vb*{S}_j
= S_i^{y_{\rm s}}S_j^{y_{\rm s}} +\eta_{ij} \left(S_i^{x_{\rm s}}S_j^{x_{\rm s}} +S_i^{z_{\rm s}}S_j^{z_{\rm s}} \right),
\end{equation}
where $\eta_{ij}=+1$ ($-1$) for an FM (AFM) bond.
For the orbital sector, we define
\begin{equation}
\tau_i^{z_{\rm o}} =\cos\alpha_i\,\tau_i^x+\sin\alpha_i\,\tau_i^y, \quad 
\tau_i^{x_{\rm o}} =-\sin\alpha_i\,\tau_i^x+\cos\alpha_i\,\tau_i^y,
\end{equation}
where $\alpha_i$ denotes the orbital pseudospin angle measured from the positive $\tau^x$ axis in the $\tau^x$-$\tau^y$ plane at site $i$. 
The bond-dependent orbital operator becomes
\begin{equation}
\tilde{\tau}_i^\gamma = c_i^\gamma\tau_i^{z_{\rm o}} +s_i^\gamma\tau_i^{x_{\rm o}},
\end{equation}
with $c_i^\gamma=\cos(\phi_\gamma-\alpha_i)$, and $s_i^\gamma=\sin(\phi_\gamma-\alpha_i)$, where $\phi_x=0$, $\phi_y=-2\pi/3$, and $\phi_z=2\pi/3$.

We introduce the local four-state basis
\begin{equation}
\begin{split}
\ket*{0_i} =\ket*{\uparrow_i}_{\rm s} \otimes \ket*{\uparrow_i}_{\rm o}, \quad 
\ket*{s_i} =\ket*{\downarrow_i}_{\rm s} \otimes \ket*{\uparrow_i}_{\rm o}, \quad
\ket*{o_i} =\ket*{\uparrow_i}_{\rm s} \otimes \ket*{\downarrow_i}_{\rm o}, \quad 
\ket*{so_i} =\ket*{\downarrow_i}_{\rm s} \otimes \ket*{\downarrow_i}_{\rm o},
\end{split}
\end{equation}
where the arrows denote the eigenstates of $S_i^{z_{\rm s}}$ and $\tau_i^{z_{\rm o}}$, respectively.
The reference state $\ket*{0_i}$ is taken as the local vacuum, while the other three states are represented by the flavor-wave bosons $b_{i,s}$, $b_{i,o}$, and $b_{i,so}$.

The local spin, orbital, and spin-orbital operators can be expressed in terms of these bosons. 
For example, their expansions to quadratic order are
\begin{equation}
\begin{split}
&S_i^{z_{\rm s}} =\frac12-b_{i,s}^\dagger b_{i,s} -b_{i,so}^\dagger b_{i,so}, \quad 
\tau_i^{x_{\rm o}} =\frac12\left( b_{i,o}^\dagger+b_{i,o} +b_{i,so}^\dagger b_{i,s} +b_{i,s}^\dagger b_{i,so} \right), \\
&S_i^{x_{\rm s}}\tau_i^{x_{\rm o}} =\frac14\left(b_{i,so}^\dagger+b_{i,so} +b_{i,s}^\dagger b_{i,o} +b_{i,o}^\dagger b_{i,s} \right).
\end{split}
\end{equation}
Here and below, terms of cubic and higher order in the bosons are omitted.
In particular, a simultaneous spin and orbital flip is represented by the single local spin-orbital flavor $b_{i,so}^\dagger$, rather than by a product $b_{i,s}^\dagger b_{i,o}^\dagger$.
Substituting these expressions into the Hamiltonian gives
\begin{equation}
\mathcal{H} = E_{\rm cl}+\mathcal{H}^{(1)}+\mathcal{H}^{(2)}+O(b^3),
\end{equation}
where $\mathcal{H}^{(n)}$ contains terms of $n$th order in the bosonic operators.

We derive the quadratic contributions from a nearest-neighbor bond $\langle ij\rangle$.
We first consider the spin exchange term, $h_{ij}^{(J)} = J \vb*{S}_i\cdot\vb*{S}_j$.
Using the local spin frames and the expansions introduced above, we obtain
\begin{equation}
\begin{split}
h_{ij}^{(J)} &=\frac{J\eta_{ij}}{4} -\frac{J\eta_{ij}}{2} (n_{i,s}+n_{j,s}+n_{i,so}+n_{j,so}) \\
&+\frac{J(1+\eta_{ij})}{4} \left( b_{i,s}^\dagger b_{j,s} +\mathrm{H.c.} \right) 
+\frac{J(\eta_{ij}-1)}{4} \left(b_{i,s}^\dagger b_{j,s}^\dagger +\mathrm{H.c.}\right) +O(b^3),
\end{split}
\end{equation}
where $n_{i,\nu}=b_{i,\nu}^\dagger b_{i,\nu}$ for $\nu=s,o,so$.
The longitudinal spin components generate onsite energies for both spin and spin-orbital fluctuations, since either excitation reverses the local spin.
The transverse components generate intersite spin hopping on FM bonds ($\eta_{ij}=+1$) and spin pairing on AFM bonds ($\eta_{ij}=-1$).

We next expand the $\kappa$ term, $h_{ij}^{(\kappa)} = 2\kappa (\tilde{\tau}_i^\gamma+\tilde{\tau}_j^\gamma) \vb*{S}_i\cdot\vb*{S}_j$.
Using $\tilde{\tau}_i^\gamma=c_i^\gamma\tau_i^{z_{\rm o}}+s_i^\gamma\tau_i^{x_{\rm o}}$, we separate the longitudinal and transverse orbital contributions: 
\begin{equation}
h_{ij}^{(\kappa,\parallel)} = 2\kappa (c_i^\gamma\tau_i^{z_{\rm o}} +c_j^\gamma\tau_j^{z_{\rm o}}) \vb*{S}_i\cdot\vb*{S}_j, \quad 
h_{ij}^{(\kappa,\perp)} = 2\kappa (s_i^\gamma\tau_i^{x_{\rm o}} +s_j^\gamma\tau_j^{x_{\rm o}}) \vb*{S}_i\cdot\vb*{S}_j. 
\end{equation}

The expansion of the longitudinal contribution gives
\begin{equation}
\begin{split}
h_{ij}^{(\kappa,\parallel)} &=
\frac{\kappa\eta_{ij}}{4} (c_i^\gamma+c_j^\gamma)
-\frac{\kappa\eta_{ij}}{2} (c_i^\gamma+c_j^\gamma) (n_{i,s}+n_{j,s})
-\frac{\kappa\eta_{ij}}{2} (c_i^\gamma n_{i,o} +c_j^\gamma n_{j,o} +c_j^\gamma n_{i,so} +c_i^\gamma n_{j,so}) \\[1mm]
&\quad +\frac{\kappa(1+\eta_{ij})}{4} (c_i^\gamma+c_j^\gamma) \left(b_{i,s}^\dagger b_{j,s} +\mathrm{H.c.} \right)
+\frac{\kappa(\eta_{ij}-1)}{4} (c_i^\gamma+c_j^\gamma)
\left(b_{i,s}^\dagger b_{j,s}^\dagger +\mathrm{H.c.} \right) +O(b^3).
\end{split}
\end{equation}
The spin hopping and pairing terms have the same structure as those generated by $h_{ij}^{(J)}$, with the exchange coefficient modified by the orbital configuration.
The orbital sector contains only onsite terms because the Hamiltonian is linear in the orbital operators.

The expansion of the transverse contribution is
\begin{equation}
\begin{split}
h_{ij}^{(\kappa,\perp)} &= 
\frac{\kappa\eta_{ij}}{4}
\left[s_i^\gamma(b_{i,o}^\dagger+b_{i,o}) +s_j^\gamma(b_{j,o}^\dagger+b_{j,o}) \right]
-\frac{\kappa\eta_{ij}}{4} \left[ s_i^\gamma(b_{i,s}^\dagger b_{i,so}+\mathrm{H.c.}) +s_j^\gamma(b_{j,s}^\dagger b_{j,so}+\mathrm{H.c.}) \right] \\[1mm]
&\quad +\frac{\kappa\eta_{ij}}{2} \left[s_j^\gamma b_{i,s}^\dagger b_{j,so} +s_i^\gamma b_{i,so}^\dagger b_{j,s} +\mathrm{H.c.} \right] +O(b^3).
\end{split}
\end{equation}
The first term describes linear orbital fluctuations, while the remaining terms hybridize spin and spin-orbital fluctuations.
For the dimer configurations, the orbital pseudospins on each FM bond satisfy $\alpha_i=\alpha_j=\phi_\gamma+\pi$, giving $s_i^\gamma=s_j^\gamma=0$.
Consequently, the hybridization terms involving boson hopping vanish, leaving only those involving pair creation and annihilation on AFM bonds.

Combining the constant terms yields the classical energy
\begin{equation}
E_{\rm cl} = \frac{1}{4} \sum_{\langle ij\rangle} \eta_{ij} \big[J +\kappa(c_i^\gamma +c_j^\gamma) \big].
\end{equation}
The linear terms are
\begin{equation}
\mathcal{H}^{(1)} =  \frac{\kappa}{4} \sum_i
\left( \sum_\gamma \eta_i^\gamma s_i^\gamma \right) (b_{i,o}^\dagger+b_{i,o}),
\end{equation}
where $\eta_i^\gamma=\eta_{ij}$, with $j$ denoting the nearest neighbor of site $i$ along the $\gamma$-bond.
For the dimer configurations, $\sum_\gamma\eta_i^\gamma s_i^\gamma=0$ holds at every site, so the linear terms vanish, consistent with the stationarity condition.
The onsite hybridization between spin and spin-orbital fluctuations is proportional to the same sum and consequently vanishes as well.

The quadratic Hamiltonian takes the form
\begin{equation}
\begin{split}
\mathcal{H}^{(2)}
&= \sum_i \left(\epsilon_s b_{i,s}^{\dagger}b_{i,s} +\epsilon_o b_{i,o}^{\dagger}b_{i,o} +\epsilon_{so} b_{i,so}^{\dagger}b_{i,so} \right) +t\sum_{\langle ij\rangle_{\rm FM}} \left(b_{i,s}^{\dagger}b_{j,s}+\mathrm{H.c.}\right) \\
&\quad +\Delta\sum_{\langle ij\rangle_{\rm AFM}} \left(b_{i,s}^{\dagger}b_{j,s}^{\dagger}+\mathrm{H.c.}\right)
-\frac{\kappa}{2}\sum_{\langle ij\rangle_{\rm AFM}} \Big[s_j^\gamma b_{i,s}^{\dagger}b_{j,so}^{\dagger} +s_i^\gamma b_{i,so}^{\dagger}b_{j,s}^{\dagger} +\mathrm{H.c.} \Big]. 
\end{split}
\end{equation}
The onsite energies are independent of the site and the dimer configuration.
This follows from the fact that each site belongs to one FM bond and two AFM bonds, for which the orbital coefficients satisfy $c_i^\gamma=-1$ and $c_i^\gamma=1/2$, respectively.
Consequently,
\begin{equation}
\sum_\gamma \eta_i^\gamma=-1,\quad
\sum_\gamma \eta_i^\gamma c_i^\gamma =\sum_\gamma \eta_i^\gamma c_j^\gamma=-2, 
\end{equation}
where $j$ denotes the nearest neighbor of site $i$ along the $\gamma$-bond.
Using these relations, we obtain
\begin{equation}
\begin{split}
\epsilon_{i,s} &= -\frac{1}{2} \sum_\gamma \eta_i^\gamma \left[ J+\kappa(c_i^\gamma +c_j^\gamma) \right] = \frac{J}{2}+2\kappa, \quad
\epsilon_{i,o} = -\frac{\kappa}{2} \sum_\gamma  \eta_i^\gamma c_i^\gamma = \kappa, \\
\epsilon_{i,so} &=  -\frac{1}{2} \sum_\gamma \eta_i^\gamma \left[J+\kappa c_j^\gamma \right] = \frac{J}{2}+\kappa.
\end{split}
\end{equation}
We henceforth omit the site index and denote them by $\epsilon_s$, $\epsilon_o$, and $\epsilon_{so}$.
The spin hopping and pairing amplitudes on FM and AFM bonds, respectively, are
\begin{equation}
t=\frac{J}{2} +\frac{\kappa}{2} (c_i^\gamma+c_j^\gamma) = \frac{J-2\kappa}{2}, \quad 
\Delta = -\frac{J}{2} -\frac{\kappa}{2} (c_i^\gamma+c_j^\gamma)=-\frac{J+\kappa}{2}.
\end{equation}

\subsection{Uniform-dimer configuration}

We now consider the uniform-dimer configuration, in which all dimers occupy the $x$-bonds.
The spins are parallel within each dimer and antiparallel across the remaining bonds, giving
\begin{equation}
\eta_x=+1,\quad \eta_y=\eta_z=-1.
\end{equation}
The orbital pseudospins are uniformly oriented along the negative $\tau^x$-axis, corresponding to $\alpha_i=\pi$ at every site.
The orbital coefficients are therefore
\begin{equation}
(c^x,c^y,c^z) =\left(-1,\frac12,\frac12\right),\quad
(s^x,s^y,s^z) =\left(0,\frac{\sqrt3}{2},-\frac{\sqrt3}{2}\right).
\end{equation}
For this configuration, the coupling between spin and spin-orbital fluctuations becomes
\begin{equation}
\begin{split}
&\quad -\frac{\kappa}{2}\sum_{\langle ij\rangle_{\rm AFM}} \Big[s_j^\gamma b_{i,s}^{\dagger}b_{j,so}^{\dagger} +s_i^\gamma b_{i,so}^{\dagger}b_{j,s}^{\dagger} +\mathrm{H.c.}\Big] \\
&=  \chi \left\{\sum_{\langle ij\rangle_y} \left(b_{i,s}^{\dagger}b_{j,so}^{\dagger} +b_{i,so}^{\dagger}b_{j,s}^{\dagger} \right) 
-\sum_{\langle ij\rangle_z} \left(b_{i,s}^{\dagger}b_{j,so}^{\dagger} +b_{i,so}^{\dagger}b_{j,s}^{\dagger} \right) \right\} +\mathrm{H.c.},
\end{split}
\end{equation}
where $\chi=-\sqrt{3}\kappa/4$. 

\begin{figure}[tbp]
	\begin{center}
		\includegraphics[width=6cm]{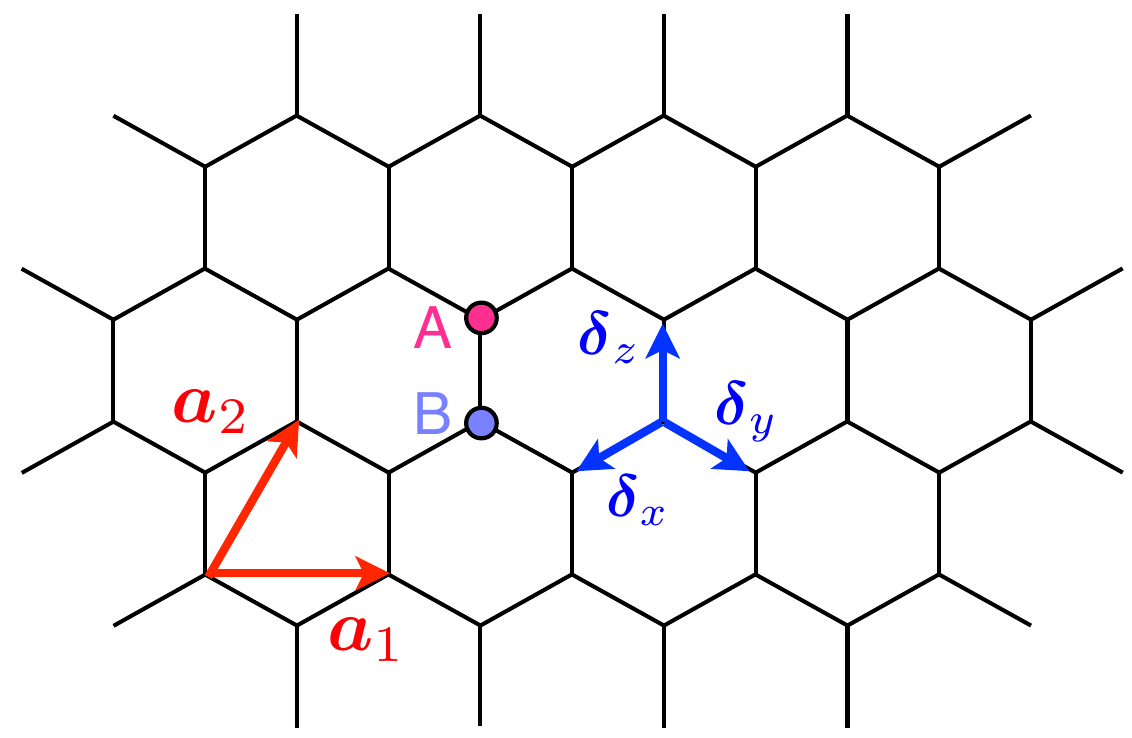}
		\caption{
			\textbf{Honeycomb lattice. }
			The two sublattices are denoted by A and B.
			The primitive translation vectors are $\vb*{a}_1$ and $\vb*{a}_2$, and the nearest-neighbor bond vectors $\vb*{\delta}_{\gamma}$ ($\gamma=x,y,z$) are oriented from the B sublattice to the A sublattice.
		}
		\label{fig:supple_lattice}
	\end{center}
\end{figure}

For the uniform-dimer configuration, the quadratic Hamiltonian is invariant under the primitive translations of the honeycomb lattice and can therefore be expressed using a two-site unit cell with sublattices A and B.
We choose the primitive translation vectors as
\begin{equation}
\vb*{a}_1=(1,0), \quad
\vb*{a}_2=\left(\frac12,\frac{\sqrt3}{2}\right),
\end{equation}
where their length is set to unity (see Fig.~\ref{fig:supple_lattice}).
The sublattice positions within the unit cell are
\begin{equation}
\vb*{d}_{\rm A}=\left(0,\frac1{\sqrt3}\right),\quad
\vb*{d}_{\rm B}=(0,0).
\end{equation}
We define the Fourier transform by
\begin{equation}
b_{\vb*{R}, \mu, \nu} = \frac{1}{\sqrt{N_c}} \sum_{\vb*{q}} e^{+i\vb*{q} \cdot(\vb*{R} +\vb*{d}_\mu)}\, b_{\vb*{q},\mu, \nu}, 
\end{equation}
where $\vb*{R}$ is a Bravais lattice vector, $\mu={\rm A}, {\rm B}$ labels the sublattice, $\nu=s,o,so$ denotes the boson flavor, and $N_c$ is the number of unit cells.
The nearest-neighbor bond vectors, directed from sublattice B to sublattice A, are
\begin{equation}
\begin{split}
\vb*{\delta}_x = -\vb*{a}_2+\vb*{d}_{\rm A} - \vb*{d}_{\rm B}, \qquad 
\vb*{\delta}_y =
\vb*{a}_1 -\vb*{a}_2 +\vb*{d}_{\rm A} -\vb*{d}_{\rm B}, \qquad 
\vb*{\delta}_z = \vb*{d}_{\rm A} -\vb*{d}_{\rm B}.
\end{split}
\end{equation}

For each momentum, we collect the flavor-wave bosons into the vector
\begin{equation}
\vb*{b}_{\vb*{q}}
= \left( b_{\vb*{q},{\rm A},s}, b_{\vb*{q},{\rm B},s}, b_{\vb*{q},{\rm A},o}, b_{\vb*{q},{\rm B},o}, b_{\vb*{q},{\rm A},so}, b_{\vb*{q},{\rm B},so} \right)^T. 
\end{equation}
The quadratic Hamiltonian is written as
\begin{equation}
\mathcal{H}^{(2)} = \sum_{\vb*{q}}
\left[ \vb*{b}_{\vb*{q}}^{\dagger}\, A(\vb*{q})\, \vb*{b}_{\vb*{q}}
+\frac{1}{2} \left( \vb*{b}_{\vb*{q}}^{\dagger}\, B(\vb*{q})\, \vb*{b}_{-\vb*{q}}^{\dagger T} + \mathrm{H.c.} \right) \right]. 
\end{equation}
For convenience, we introduce the momentum-dependent functions
\begin{equation}
h(\vb*{q}) = e^{i\vb*{q}\cdot\vb*{\delta}_x}, \quad 
f(\vb*{q}) = e^{i\vb*{q}\cdot\vb*{\delta}_y} + e^{i\vb*{q}\cdot\vb*{\delta}_z}, \quad 
g(\vb*{q}) = e^{i\vb*{q}\cdot\vb*{\delta}_y} -e^{i\vb*{q}\cdot\vb*{\delta}_z}.
\end{equation}
In the basis defined above, the matrix $A(\vb*{q})$ takes the block-diagonal form
\begin{equation}
\begin{split}
& A(\vb*{q}) = \mqty(A_s(\vb*{q}) & 0 & 0 \\ 0 & A_o(\vb*{q}) & 0 \\ 0 & 0 & A_{so}(\vb*{q})), \\[2mm] 
& A_s(\vb*{q}) = \mqty( \epsilon_s & t h^*(\vb*{q}) \\ t h(\vb*{q}) & \epsilon_s ), \quad 
A_o(\vb*{q}) = \mqty(\epsilon_o & 0 \\ 0 & \epsilon_o), \quad 
A_{so}(\vb*{q}) = \mqty(\epsilon_{so} & 0 \\ 0 & \epsilon_{so}). 
\end{split}
\end{equation}
The matrix $B(\vb*{q})$, describing pair creation and annihilation, is
\begin{equation}
\begin{split}
B(\vb*{q}) &= \mqty( B_s(\vb*{q}) & 0 & B_{s,so}(\vb*{q}) \\ 0 & 0 & 0 \\ B_{so,s}(\vb*{q}) & 0 & 0 ), \\[2mm]
B_s(\vb*{q}) &=\mqty( 0 & \Delta f^*(\vb*{q}) \\ \Delta f(\vb*{q}) & 0), \quad
B_{s,so}(\vb*{q}) = B_{so,s}(\vb*{q}) = \mqty( 0 & \chi g^*(\vb*{q}) \\ \chi g(\vb*{q}) & 0 ).
\end{split}
\end{equation}
Introducing the Nambu spinor
\begin{equation}
\Phi_{\vb*{q}} = \mqty(\vb*{b}_{\vb*{q}}\\ \vb*{b}_{-\vb*{q}}^\dagger ),
\end{equation}
we obtain the Bogoliubov-de Gennes Hamiltonian
\begin{equation}
\mathcal{H}^{(2)} = \frac{1}{2} \sum_{\vb*{q}} \Phi_{\vb*{q}}^{\dagger}\, H_{\rm BdG}(\vb*{q})\, \Phi_{\vb*{q}} 
-\frac{1}{2} \sum_{\vb*{q}} \Tr A(\vb*{q}),
\end{equation}
where
\begin{equation}
H_{\rm BdG}(\vb*{q}) = \mqty( A(\vb*{q}) & B(\vb*{q}) \\ B^\dagger(\vb*{q}) & A^{\rm T}(-\vb*{q}) ). 
\end{equation}
The excitation energies are obtained from the eigenvalues of the $12\times12$ dynamical matrix
\begin{equation}
D(\vb*{q}) =  \Sigma_z H_{\rm BdG}(\vb*{q}), \quad
\Sigma_z= \mqty(I_6 & 0 \\ 0 & -I_6).
\end{equation}
The orbital sector is decoupled from the spin and spin-orbital sectors and gives two flat bands with energy $\epsilon_o$.
The remaining four positive-energy bands, denoted by $\omega_n(\vb*{q})$ with $n=1,\ldots,4$, belong to the spin and spin-orbital sectors, which are hybridized by the $\kappa$ interaction. 
At the $\Gamma$ point, the hybridization between spin and spin-orbital fluctuations vanishes because $g(\vb*{0})=0$.
The spin sector contains a gapless mode, corresponding to the Goldstone mode associated with the spontaneous breaking of spin-rotation symmetry.

\subsection{Goldstone-mode instability}

We now examine the long-wavelength instability of the Goldstone mode in the uniform-dimer configuration.
We decompose the momentum as $\vb*{q} = q_{\parallel} \vb*{e}_{\parallel} +q_{\perp}\vb*{e}_{\perp}$, where
\begin{equation}
\vb*{e}_{\parallel} =  \left(\frac{\sqrt3}{2},\frac12\right), \quad
\vb*{e}_{\perp} =  \left(-\frac12,\frac{\sqrt3}{2}\right)
\end{equation}
are parallel and perpendicular to the dimers, respectively.
The direction $\vb*{e}_{\perp}$ corresponds to the $\Gamma$-$K$ direction, along which the Goldstone mode becomes unstable.

Since the orbital sector is decoupled, we restrict the dynamical matrix to the spin and spin-orbital sectors, yielding an $8\times8$ matrix.
Along this direction, $q_{\parallel}=0$, the momentum-dependent functions introduced above reduce to
\begin{equation}
h(q_{\perp})=1,\quad
f(q_{\perp})=2 c_q,\quad
g(q_{\perp})=-2i s_q, 
\end{equation}
where $c_q=\cos(q_{\perp}/2)$ and $s_q=\sin(q_{\perp}/2)$.
The corresponding sublattice structure is
\begin{equation}
A_s(q_{\perp}) =\epsilon_s\sigma_0+t\sigma_x, \quad
B_s(q_{\perp}) =2\Delta c_q\,\sigma_x, \quad
B_{s,so}(q_{\perp}) =-2\chi s_q\,\sigma_y,
\end{equation}
where $\sigma_\mu$ are Pauli matrices acting in the sublattice space.
We introduce the bonding and antibonding combinations
\begin{equation}
b_{\vb*{q},\pm,\nu} = \frac{b_{\vb*{q},{\rm A},\nu}\pm b_{\vb*{q},{\rm B},\nu}}{\sqrt2}, \quad \nu=s,so.
\end{equation}
Since the hybridization is proportional to $\sigma_y$, it couples opposite sublattice parities, $s_+\leftrightarrow so_-$ and $s_-\leftrightarrow so_+$.
Consequently, the $8\times8$ dynamical matrix separates into two independent $4\times4$ blocks, $D(q_{\perp})=D_+(q_{\perp})\oplus D_-(q_{\perp})$. 
The block $D_+(q_{\perp})$, which contains the Goldstone mode, is given by
\begin{equation}
D_+(q_{\perp}) = 
\mqty(
\epsilon_s+t & 0 & 2\Delta c_q & 2i\chi s_q \\
0 & \epsilon_{so} & 2i\chi s_q & 0 \\
-2\Delta c_q & 2i\chi s_q & -(\epsilon_s+t) & 0 \\
2i\chi s_q & 0 & 0 & -\epsilon_{so}). 
\end{equation}
To obtain the low-energy dispersion, we eliminate the gapped spin-orbital components from the eigenvalue equation
\begin{equation}
D_+(q_{\perp}) \mqty(u_s \\ u_{so} \\ v_s \\ v_{so}) =  \varepsilon \mqty(u_s \\ u_{so} \\ v_s \\ v_{so}).
\end{equation}
This yields an effective eigenvalue problem for the spin components,
\begin{equation}
\mqty(\epsilon_s +t -\varepsilon-\Sigma_+ (\varepsilon, q_{\perp}) & 2\Delta c_q \\[1mm]
2\Delta c_q & \epsilon_s +t+\varepsilon-\Sigma_- (\varepsilon, q_{\perp}) ) \mqty(u_s \\ v_s)=0,
\end{equation}
where
\begin{equation}
\Sigma_{\pm}(\varepsilon,q_{\perp}) =  \frac{4\chi^2 s_q^2}{\epsilon_{so}\pm\varepsilon}.
\end{equation}
These terms represent the correction to the spin-sector dynamics generated by virtual fluctuations of the gapped spin-orbital mode.

To determine the leading long-wavelength dispersion of the Goldstone mode, we expand the effective eigenvalue equation in small $q_\perp$ and $\varepsilon$: 
\begin{equation}
\begin{split}
4\chi^2 s_q^2 =\chi^2 q_{\perp}^2+O(q_{\perp}^4), \quad 
2\Delta c_q =2\Delta\left(1-\frac{q_{\perp}^2}{8}\right)+O(q_{\perp}^4), \quad 
\frac{1}{\epsilon_{so}\pm\varepsilon} = \frac{1}{\epsilon_{so}}+O(q_{\perp}).  
\end{split}
\end{equation}
Substituting these expansions into the determinant of the effective eigenvalue problem, we obtain
\begin{equation}
\begin{split}
0 =(\epsilon_s+t)^2-4\Delta^2-\varepsilon^2 +\left[\Delta^2 -\frac{2(\epsilon_s+t)\chi^2}{\epsilon_{so}} \right]q_{\perp}^2 +O(q_{\perp}^4).
\end{split}
\end{equation}
Using the condition $(\epsilon_s+t)^2=4\Delta^2$, we find
\begin{equation}
\varepsilon_G^2(q_{\perp}) =  \rho_{\perp}q_{\perp}^2 +O(q_{\perp}^4),
\end{equation}
where
\begin{equation}\label{eq:goldstone_rho_perp}
\rho_{\perp} = \Delta^2 -\frac{2(\epsilon_s+t)\chi^2}{\epsilon_{so}} = \frac{(J+\kappa)^2}{4} - \frac{3\kappa^2(J+\kappa)}{8(J/2+\kappa)}.
\end{equation}
The first term includes the stiffness of the spin mode in the absence of hybridization, while the second term represents the negative correction arising from its coupling to the gapped spin-orbital mode.

The long-wavelength instability occurs when $\rho_{\perp}=0$, which gives $J^2+3J\kappa-\kappa^2=0$. 
The corresponding critical coupling is
\begin{equation}\label{eq:goldstone_kappa_c}
\frac{\kappa_c}{J} = \frac{3+\sqrt{13}}2 \simeq 3.303.
\end{equation}
At this coupling, the stiffness perpendicular to the dimers vanishes, while the stiffness in the parallel direction remains positive.

To determine the dispersion at the critical point, we retain the next-order contribution,
\begin{equation}
\varepsilon_G^2(q_{\perp}) =  \rho_{\perp}q_{\perp}^2 +\beta_{\perp}q_{\perp}^4 +O(q_{\perp}^6).
\end{equation}
At $\kappa=\kappa_c$, the quartic coefficient is
\begin{equation}
\beta_{\perp,c} = \frac{9\kappa_c^4}{256(J/2+\kappa_c)^2} = \frac{(J+\kappa_c)^2}{64}>0.
\end{equation}
Consequently, the Goldstone mode exhibits a quadratic dispersion along the perpendicular direction at the critical coupling,
\begin{equation}
\omega_G(q_{\perp}) = \frac{J+\kappa_c}{8}\, q_{\perp}^2 +O(q_{\perp}^4).
\end{equation}

Finally, we relate the flavor-wave analysis to the effective Lagrangian introduced in the main text.
For $\vb*{q}=q_\parallel\vb*{e}_\parallel+q_\perp\vb*{e}_\perp$, the momentum dependence of the hybridization between spin and spin-orbital fluctuations is given by
\begin{equation}
g(q_\parallel,q_\perp)
=-2i e^{iq_\parallel/(2\sqrt3)} \sin(q_\perp/2)
=-iq_\perp+\frac{q_\parallel q_\perp}{2\sqrt3}+O(q^3).
\end{equation}
Thus, the hybridization vanishes at the $\Gamma$ point and is linear in $q_\perp$ to leading order, with no term linear in $q_\parallel$.
In the continuum description, this leading momentum dependence gives rise to a spatial derivative coupling between the Goldstone field $\theta$ and a gapped spin-orbital field $\phi$, of the form $\phi\,\partial_\perp\theta$.

At energies well below the spin-orbital gap, integrating out $\phi$ generates a negative correction to the coefficient of $(\partial_\perp\theta)^2$.
In the phenomenological Lagrangian of the main text, this correction takes the form $-\gamma^2/\Delta_{so}^2$.
Its microscopic counterpart is the second term in Eq.~\eqref{eq:goldstone_rho_perp},
\begin{equation}
\delta\rho_\perp = -\frac{2(\epsilon_s+t)\chi^2}{\epsilon_{so}}.
\end{equation}

\section{Finite-size analysis of the incommensurate spiral state}

The incommensurate spiral state is characterized by a modulation wave vector that is incompatible with the periodicity of a finite cluster.
Under periodic boundary conditions, the allowed wave vectors are discrete, making a finite-size analysis necessary to estimate the modulation wave vector in the thermodynamic limit.
In this section, we describe the finite-size analysis of the modulation wave vector, examine the coplanarity of the optimized spin configurations, and explain the procedure used to evaluate the local spin-orbital entanglement and ordered moments.

\subsection{Modulation wave vector}

We consider $L\times L$ honeycomb clusters with periodic boundary conditions, where $L=56$, 64, 72, 80, 88, 96, 104, 112, and 120. 
For each value of $\kappa$, we include only the system sizes for which the numerical optimization converged. 

To characterize the incommensurate spiral state, we calculate the spin structure factor $S_{\rm spin}(\vb*{q})$. 
The dominant peaks of the spiral state occur at
\begin{equation}
\vb*{Q}_{\pm} = \vb*{Q}_0 \pm q_L\vb*{e}_\perp,
\end{equation}
where $\vb*{Q}_0$ is the ordering wave vector of the uniform-dimer configuration and $\vb*{e}_\perp$ is the unit vector along the modulation direction.
We choose the extended-zone representative
\begin{equation}
\vb*{Q}_0 = \frac{-\vb*{b}_1+\vb*{b}_2}{2} = \left(-\pi, \sqrt{3}\pi\right),
\end{equation}
with the reciprocal lattice vectors
\begin{equation}
\vb*{b}_1 = \left(2\pi,-\frac{2\pi}{\sqrt{3}} \right), \quad
\vb*{b}_2 = \left(0,\frac{4\pi}{\sqrt{3}} \right).
\end{equation}

On an $L\times L$ cluster, the allowed modulation wave numbers are
\begin{equation}
q_{m,L} = \frac{4\pi |m|}{L}, \quad m\in\mathbb{Z}.
\end{equation}
For each optimized state, we identify the integer $m_L$ corresponding to the dominant pair of peaks in $S_{\rm spin}(\vb*{q})$.
The finite-size modulation wave number is then $q_L = 4\pi m_L/L$. 
Because the allowed momenta are discrete, $q_L$ does not vary continuously with $L$.
Instead, the changes in $m_L$ lead to nonmonotonic finite-size variations of $q_L$. 

To estimate the modulation wave number $q_*$, we assume that the wave number selected on a finite cluster is the nearest allowed value to $q_*$.
Under this assumption, the observed integer $m_L$ constrains $q_*$ to
\begin{equation}
\frac{4\pi}{L} \left(m_L-\frac12\right) \leq
q_* \leq \frac{4\pi}{L} \left(m_L+\frac12\right).
\end{equation}
Combining these constraints over the converged system sizes at each $\kappa$, we obtain the bounds
\begin{equation}
q_*^{\rm min} = \max_L \left[ \frac{4\pi}{L} \left(m_L-\frac12\right) \right], \quad 
q_*^{\rm max} = \min_L \left[ \frac{4\pi}{L} \left(m_L+\frac12\right) \right].
\end{equation}
Here, the maximum and minimum are taken only over the system sizes for which the numerical optimization converged.

When the resulting intervals have a nonempty intersection, the finite-size data are consistent with a common wave number within the nearest-grid approximation.
We take the midpoint of this interval,
\begin{equation}
q_* =  \frac{q_*^{\rm min}+q_*^{\rm max}}{2},
\end{equation}
as our estimate of the modulation wave number in the thermodynamic limit.
The interval $[q_*^{\rm min}, q_*^{\rm max}]$ represents the range compatible with the finite-size momentum quantization rather than a statistical uncertainty.

\subsection{Coplanarity of the magnetic structure}

To characterize the magnetic structure of the optimized spiral state, we examine the relative orientations of the ordered spins. 
For all model parameters and system sizes considered, the two spins within each dimer remain collinear and parallel, allowing us to define a dimer spin $\vb*{D}_n$. 
Furthermore, the dimer spins are ferromagnetically aligned along the direction parallel to the dimers, $\vb*{e}_\parallel$, without any modulation. 
We therefore focus on the variation of the dimer spins along the perpendicular direction $\vb*{e}_\perp$, labeling successive dimers by $n$.

To quantify the coplanarity of the spin configuration, we introduce
\begin{equation}
C = \sqrt{1-\left\lvert \frac{1}{L} \sum_n \frac{\vb*{D}_n \cross \vb*{D}_{n+1}}{|\vb*{D}_n \cross \vb*{D}_{n+1}|} \right\rvert^2}. 
\end{equation}
The normalized cross product defines the normal to the plane spanned by neighboring dimer spins. 
For a coplanar spiral with a uniform sense of rotation, these normals are aligned, giving $C=0$. 
In contrast, a noncoplanar configuration exhibits spatial variations of the local normal directions, resulting in a finite $C$.

\begin{figure*}[t]
	\centering
	\includegraphics[width=9cm]{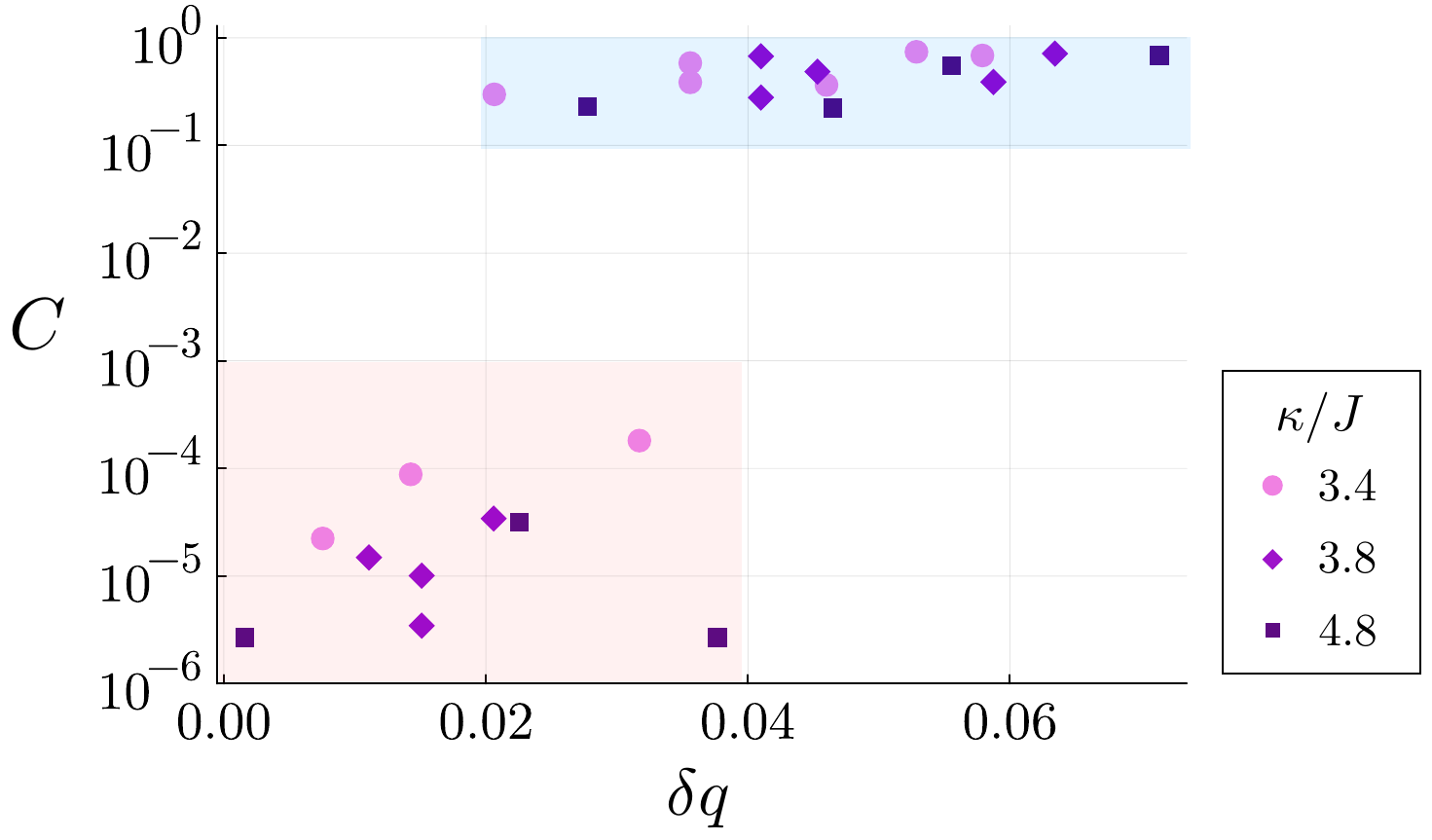}
	\caption{
		\textbf{Coplanarity of the optimized spiral state.}
		Coplanarity measure $C$ as a function of the wave-number mismatch $\delta q=|q_*-q_L|$ for $\kappa/J=3.4$, $3.8$, and $4.8$, obtained from finite-size clusters.
		Here, $q_L$ is the modulation wave number of the optimized configuration on an $L\times L$ cluster, and $q_*$ is its estimated value in the thermodynamic limit.
		The red- and blue-shaded regions highlight the two groups of configurations with $C\sim10^{-6}$-$10^{-3}$ and $C\gtrsim10^{-1}$, respectively.
		The former corresponds to nearly coplanar configurations, whereas the latter exhibits substantial variations in the local rotation-plane normals.
	}
	\label{fig:supple_coplanar}
\end{figure*}

We evaluate $C$ for the numerically optimized configurations at different values of $\kappa/J$ and system sizes $L$. 
To examine its relation to finite-size momentum quantization, we compare $C$ with $\delta q=|q_*-q_L|$. 
Figure~\ref{fig:supple_coplanar} plots $C$ against $\delta q$ for $\kappa/J=3.4$, $3.8$, and $4.8$.
The results separate into two groups. 
One group exhibits very small values, $10^{-6}\lesssim C\lesssim10^{-3}$, with $0<\delta q<0.04$, indicating nearly coplanar configurations. 
The other group has substantially larger values, $C\gtrsim10^{-1}$, with $0.02<\delta q<0.08$, indicating pronounced variations in the local rotation-plane normals.
The nearly coplanar configurations are found at relatively small $\delta q$, suggesting that deviations from coplanarity may be associated with finite-size momentum quantization. 

\subsection{Local spin-orbital entanglement and ordered moments}

For each value of $\kappa/J$, we select the system size $L$ with the smallest wave-number mismatch $\delta q=|q_*-q_L|$ among the clusters for which the numerical optimization converged. 
Using the corresponding optimized variational state, we evaluate the local spin-orbital entanglement entropy $S_{\rm EE}$ and the spin and orbital ordered moments. 
The resulting values are presented in the main text.

\section{Microscopic origin of the spin-orbital model}

In this section, we discuss the microscopic origin of the spin-orbital model introduced in the main text. 
We derive the effective exchange interactions from a multiorbital Hubbard model by the strong-coupling expansion and identify the conditions under which bilinear orbital exchanges are suppressed while the $\kappa$ term remains finite. 
We then examine whether the required hopping conditions can be realized in edge-sharing octahedra.

\subsection{Strong-coupling expansion of the multiorbital Hubbard model}

We consider transition-metal ions with one or two electrons occupying the $t_{2g}$ orbitals. 
Under a trigonal crystal field, the $t_{2g}$ manifold splits into an $a_{1g}$ singlet and an $e_g^\pi$ doublet. 
When the orbital degree of freedom associated with the $e_g^\pi$ doublet remains active at low energies, the strong-coupling expansion of the multiorbital Hubbard model gives rise to an effective spin-orbital Hamiltonian.

Our starting point is
\begin{equation}
\mathcal{H} = \mathcal{H}_{\rm C} + \mathcal{H}_{\rm CEF}^{\rm tri} + \mathcal{H}_{\rm kin},
\end{equation}
where $\mathcal{H}_{\rm C}$ describes the on-site Coulomb interaction, $\mathcal{H}_{\rm CEF}^{\rm tri}$ the trigonal crystal field, and $\mathcal{H}_{\rm kin}$ the hopping between neighboring transition-metal ions.

The on-site Coulomb interaction is described by the Kanamori Hamiltonian~\cite{Kanamori1963, Georges2013},
\begin{equation}
\mathcal{H}_{\rm C} =  \sum_i \left[\frac{U-3J_{\rm H}}{2} N_i(N_i-1) -2J_{\rm H} \vb*{S}_i^2 -\frac{J_{\rm H}}{2} \vb*{L}_i^2 +\frac{5}{2} J_{\rm H} N_i \right],
\end{equation}
where $U$ and $J_{\rm H}$ are the intra-orbital Coulomb interaction and Hund's coupling, respectively, and $N_i$, $\vb*{S}_i$, and $\vb*{L}_i$ denote the total electron number, spin, and orbital angular momentum at site $i$.

The trigonal crystal field is given by
\begin{equation}
\mathcal{H}_{\rm CEF}^{\rm tri} =  \Delta_{\rm tri} \sum_i \vb*{c}_i^\dagger \left[ (\hat{\vb*{n}}_i\cdot\vb*{l})^2 \otimes\sigma^0 \right] \vb*{c}_i,
\end{equation}
where $\hat{\vb*{n}}_i$ is the local trigonal axis and $\vb*{l}$ denotes the orbital angular momentum matrices in the $t_{2g}$ manifold. 
Here, $\vb*{c}_i$ collects the electron annihilation operators for the three $t_{2g}$ orbitals and two spin states.

The trigonal crystal field separates the $a_{1g}$ singlet from the $e_g^\pi$ doublet by an energy $|\Delta_{\rm tri}|$. 
For $\Delta_{\rm tri} < 0$, the $e_g^\pi$ doublet lies below the $a_{1g}$ singlet and provides the low-energy orbital degree of freedom for the $d^1$ configuration. 
For $\Delta_{\rm tri} > 0$, the $a_{1g}$ singlet lies below the doublet, and the $d^2$ configuration can retain an active orbital degree of freedom associated with the partially occupied $e_g^\pi$ doublet.

Using the eigenstates of the effective orbital angular momentum along the trigonal axis, the corresponding orbital wave functions are
\begin{equation}
\ket*{0} = \frac{1}{\sqrt{3}} \left( \ket*{yz}+\ket*{zx}+\ket*{xy} \right), \quad
\ket*{\pm1} = \pm\frac{1}{\sqrt{3}} \left( \ket*{yz} +\omega^{\pm 1}\ket*{zx} +\omega^{\mp 1}\ket*{xy} \right),
\end{equation}
where $\omega=e^{2\pi i/3}$. 
The state $\ket*{0}$ corresponds to the $a_{1g}$ singlet, while $\ket*{\pm 1}$ span the $e_g^\pi$ doublet.

Within the $e_g^\pi$ doublet, we introduce the orbital pseudospin operators
\begin{equation}
\begin{split}
\tau^x &= \frac{1}{2} \left( \ketbra*{+1}{-1} + \ketbra*{-1}{+1} \right), \quad 
\tau^y = -\frac{i}{2} \left( \ketbra*{+1}{-1} -\ketbra*{-1}{+1} \right), \\
\tau^z &= \frac{1}{2} \left( \ketbra*{+1}{+1} -\ketbra*{-1}{-1} \right).
\end{split}
\end{equation}
The transverse components $\tau^x$ and $\tau^y$ represent orbital quadrupoles, whereas $\tau^z$ represents an orbital magnetic dipole.

The kinetic Hamiltonian is written as
\begin{equation}
\mathcal{H}_{\rm kin} = \sum_{\langle i,j\rangle} \vb*{c}_i^\dagger (T_{ij}\otimes\sigma^0) \vb*{c}_j +\mathrm{H.c.},
\end{equation}
where $T_{ij}$ is the hopping matrix in the $t_{2g}$ orbital space. 
For a representative $z$-bond, symmetry constrains the hopping matrix in the cubic-orbital basis $\{\ket*{yz},\ket*{zx},\ket*{xy}\}$ to the form
\begin{equation}\label{eq:app_hopping_cubic}
T_z =  \mqty(
t_1 & t_2 & t_4 \\
t_2 & t_1 & t_4 \\
t_4 & t_4 & t_3).
\end{equation}
The hopping matrices for the other bond directions are related to $T_z$ by threefold rotations.

To derive the effective exchange interactions, we treat $\mathcal{H}_{\rm kin}$ as a perturbation to $\mathcal{H}_0=\mathcal{H}_{\rm C}+\mathcal{H}_{\rm CEF}^{\rm tri}$.
To second order in the hopping, the effective Hamiltonian within the ground-multiplet manifold is
\begin{equation}
\mathcal{H}_{\rm ex} =  \mathcal{P} \mathcal{H}_{\rm kin} \frac{1}{E_0-\mathcal{H}_0} \mathcal{H}_{\rm kin} \mathcal{P},
\end{equation}
where $\mathcal{P}$ projects onto the low-energy manifold and $E_0$ is the unperturbed energy of this manifold.

Evaluating the virtual hopping processes gives the effective spin-orbital exchange Hamiltonian~\cite{Nakai2026}
\begin{equation}\label{eq:app_general_exchange}
\begin{split}
\mathcal{H}_{\rm ex} &= \sum_{\langle i,j \rangle} \Big[ 
K_{zz} (1+\varrho_{\parallel} \vb*{S}_i\cdot\vb*{S}_j) \tau_i^Z \tau_j^Z - K_{\pm} (1+\varrho_{\perp}\vb*{S}_i\cdot\vb*{S}_j) (\tau_i^+\tau_j^- +{\rm H.c.}) \\[-1mm]
&\quad\qquad + K_{\pm\pm} (1+\varrho_{\perp}\vb*{S}_i\cdot\vb*{S}_j) (\gamma^*_{ij}\tau_i^+\tau_j^+ +{\rm H.c.}) +J\vb*{S}_i\cdot\vb*{S}_j +\kappa \vb*{S}_i\cdot\vb*{S}_j (\gamma_{ij}[\tau_i^+ +\tau_j^+] +{\rm H.c.}) \Big]. 
\end{split}
\end{equation}
Here, $\tau^\pm=\tau^x\pm i\tau^y$, and the bond-dependent phase factor $\gamma_{ij}$ takes the values $1$, $\omega$, and $\omega^2$ for the $x$-, $y$-, and $z$-bonds, respectively. 
The coefficients $\varrho_\parallel$ and $\varrho_\perp$ characterize the spin dependence of the bilinear orbital exchanges.

The first three terms describe bilinear orbital exchanges and their spin-coupled counterparts, while the fourth term represents the Heisenberg exchange. 
The last term, proportional to $\kappa$, acts as an intrinsic orbital field. 
The minimal model studied in the main text is obtained when the bilinear orbital exchanges are suppressed, leaving the Heisenberg exchange and the $\kappa$ term.

\subsection{Suppression of bilinear orbital exchanges}

We next identify the microscopic conditions under which the bilinear orbital exchanges in Eq.~(\ref{eq:app_general_exchange}) vanish. 
To this end, we express the hopping matrix in the basis $\{\ket*{+1}, \ket*{0}, \ket*{-1}\}$:
\begin{equation}
T_z = \mqty(
t_a & \omega t_c & \omega^2 t_d \\[1mm]
\omega^2 t_c & t_b & -\omega t_c \\[1mm]
\omega t_d & -\omega^2 t_c & t_a ),
\end{equation}
where
\begin{equation}
\begin{split}
t_a &= \frac{2t_1-t_2+t_3-2t_4}{3}, \quad
t_b = \frac{2t_1+2t_2+t_3+4t_4}{3}, \\[1mm]
t_c &= \frac{-t_1-t_2+t_3+t_4}{3}, \quad
t_d = \frac{t_1-2t_2-t_3+2t_4}{3}.
\end{split}
\end{equation}
The hopping integrals $t_a$ and $t_d$ describe hopping within the $e_g^\pi$ doublet, while $t_c$ describes hopping between the $e_g^\pi$ doublet and the $a_{1g}$ singlet.

The second-order strong-coupling expansion gives the bilinear orbital exchange coefficients
\begin{equation}
\begin{split}
K_{zz} = \frac{4(t_a^2-t_d^2)} {(1+\varrho_\parallel)(U-3J_{\rm H})}, \quad
K_\pm = -\frac{2t_a^2} {(1+\varrho_\perp)(U-3J_{\rm H})}, \quad
K_{\pm\pm} = \frac{2t_d^2} {(1+\varrho_\perp)(U-3J_{\rm H})}.
\end{split}
\end{equation}
Consequently, all three bilinear orbital exchange coefficients vanish when
\begin{equation}\label{eq:app_suppression_condition}
t_a=t_d=0.
\end{equation}
This condition eliminates hopping entirely within the $e_g^\pi$ doublet.
In terms of the hopping parameters in the cubic-orbital basis, Eq.~(\ref{eq:app_suppression_condition}) is equivalent to
\begin{equation}\label{eq:app_hopping_condition}
t_1=t_2, \quad
t_4=\frac{t_2+t_3}{2}.
\end{equation}

To clarify the remaining hopping processes, we introduce the bond-adapted orbital basis
\begin{equation}
\ket*{z,+} = \frac{1}{\sqrt{2}} (\ket*{yz}-\ket*{zx}), \quad
\ket*{z,-} = \frac{1}{\sqrt{6}} (\ket*{yz}+\ket*{zx}-2\ket*{xy}).
\end{equation}
These states are eigenstates of the bond-dependent orbital pseudospin $\tilde{\tau}^z$ introduced in the main text.
In the basis $\{\ket*{z,+},\ket*{z,-},\ket*{0}\}$, the hopping matrix becomes
\begin{equation}
T_z = \mqty(
t_1-t_2 & 0 & 0 \\
0 & \dfrac{t_1+t_2+2t_3-4t_4}{3} & \dfrac{\sqrt{2}(t_1+t_2-t_3-t_4)}{3} \\
0 & \dfrac{\sqrt{2}(t_1+t_2-t_3-t_4)}{3} & \dfrac{2t_1+2t_2+t_3+4t_4}{3} ).
\end{equation}
Under the conditions in Eq.~(\ref{eq:app_hopping_condition}), this reduces to
\begin{equation}\label{eq:app_reduced_hopping}
T_z = \mqty(
0 & 0 & 0 \\
0 & 0 & \dfrac{t_2-t_3}{\sqrt{2}} \\
0 & \dfrac{t_2-t_3}{\sqrt{2}} & 2t_2+t_3 ).
\end{equation}
Thus, hopping within the $e_g^\pi$ doublet is absent, while hopping between the $\ket*{z,-}$ orbital and the $a_{1g}$ singlet remains finite. 
The suppression of the bilinear orbital exchanges therefore does not require the complete suppression of orbital-dependent hopping. 
Instead, the remaining virtual hopping processes involving the $a_{1g}$ orbital can generate the spin-orbital coupling proportional to $\kappa$.

This mechanism differs from that in conventional two-orbital models, such as those based on the $e_g$ orbitals, where hopping is restricted to the active orbital doublet. 
In such models, eliminating the intra-doublet hopping suppresses all superexchange interactions. 
In contrast, the present effective two-orbital system retains virtual hopping processes involving the $a_{1g}$ orbital, allowing the bilinear orbital exchanges to vanish while the Heisenberg exchange and the $\kappa$ term remain finite.

\begin{figure*}[t]
	\centering
	\includegraphics[width=17cm]{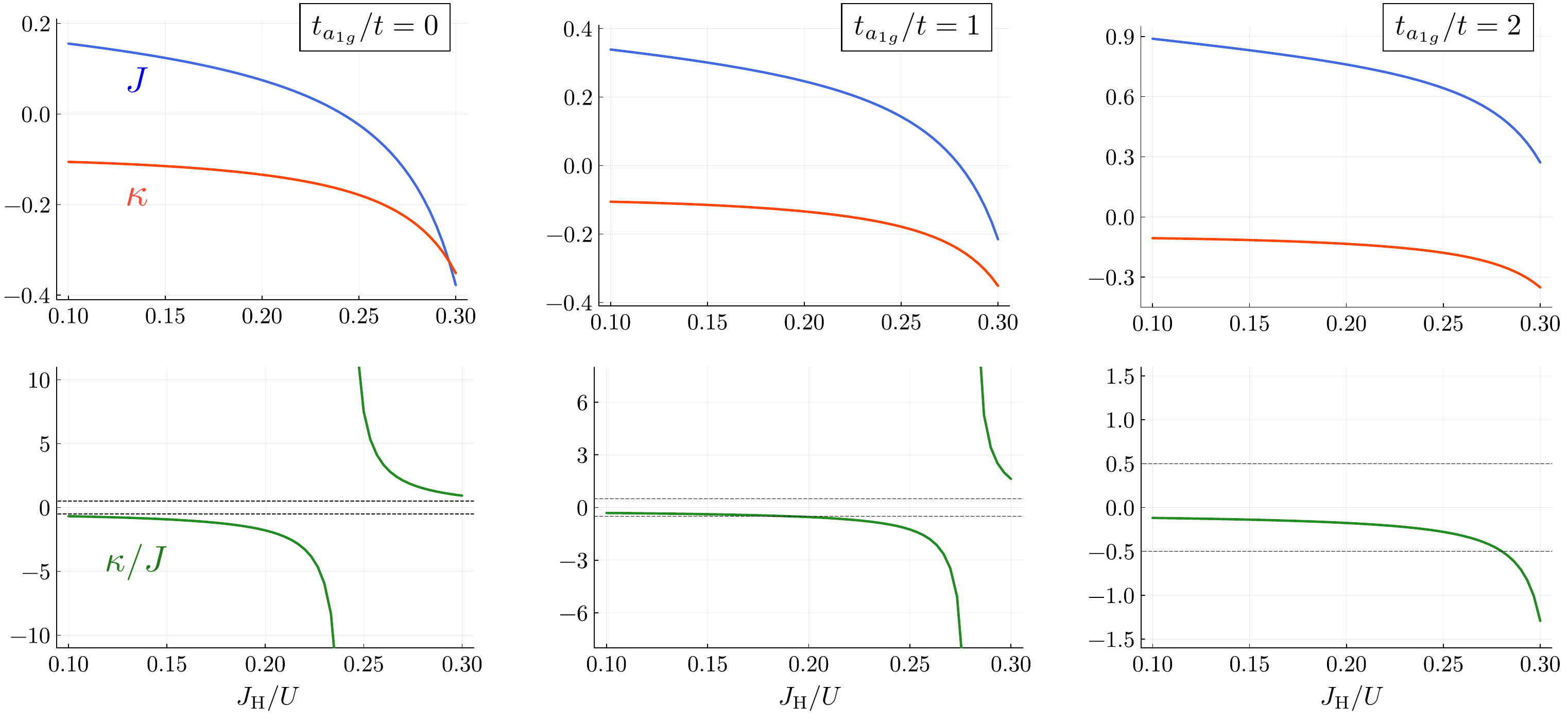}
	\caption{
		\textbf{Exchange parameters for the $d^2$ configuration.}
		Hund's coupling dependence of the exchange parameters $J$ and $\kappa$ (top) and their ratio $\kappa/J$ (bottom) for $t_{a_{1g}}/t=0$, $1$, and $2$ in the left, middle, and right panels, respectively.
		The dashed horizontal lines in the bottom panels indicate $\kappa/J=\pm 1/2$, marking the classical phase boundaries between the dimer configurations and the N\'eel state ($J>0$) or uniformly polarized state ($J<0$).
		The parameters are set to $U/t=5$ and $|\Delta_{\rm tri}|/t=0.05$.
	}
	\label{fig:supple_exchange}
\end{figure*}

We next examine the exchange parameters $J$ and $\kappa$ under the hopping conditions in Eq.~(\ref{eq:app_hopping_condition}) for the $d^1$ and $d^2$ configurations.
We take the hopping integral between the $\ket*{z,-}$ orbital and the $a_{1g}$ singlet, $t=(t_2-t_3)/\sqrt{2}$, as the energy unit and set $U/t=5$ and $|\Delta_{\rm tri}|/t=0.05$.
The sign of $\Delta_{\rm tri}$ is chosen according to the electronic configuration so that the $e_g^\pi$ doublet provides the active low-energy orbital degree of freedom, as discussed above.
We vary the Hund's coupling $J_{\rm H}/U$ and the dimensionless hopping ratio $t_{a_{1g}}/t$, where $t_{a_{1g}}=2t_2+t_3$ denotes the hopping integral between the $a_{1g}$ orbitals.
We have also verified that the bilinear orbital exchange coefficients vanish, $K_{zz}=K_\pm=K_{\pm\pm}=0$, throughout the parameter range considered.

For the $d^1$ configuration, we find that the exchange parameters satisfy $\kappa/J=-1/2$, independently of $J_{\rm H}/U$ and $t_{a_{1g}}/t$.
The Heisenberg exchange is always FM ($J<0$), while $\kappa$ remains positive.
Although the Heisenberg exchange has the opposite sign to that considered in the main text, the two cases are related at the level of site-product states without local spin-orbital entanglement.
Indeed, a global reversal of the orbital pseudospins, $\vb*{\tau}_i \to -\vb*{\tau}_i$, combined with a reversal of the spins on one sublattice, $\vb*{S}_i \to -\vb*{S}_i$ for $i\in\mathrm{B}$, maps the energy functional of the model with parameters $(J,\kappa)$ onto that with $(-J,\kappa)$.
Thus, the $d^1$ configuration is related to the AFM regime studied in the main text within this restricted class of states.
In particular, the corresponding ratio $\kappa/J=1/2$ coincides with the classical phase boundary between the N\'eel and dimer configurations.

For the $d^2$ configuration, the exchange parameters depend on both $J_{\rm H}/U$ and $t_{a_{1g}}/t$.
Figure~\ref{fig:supple_exchange} shows $J$, $\kappa$, and their ratio $\kappa/J$ as functions of $J_{\rm H}/U$ for $t_{a_{1g}}/t=0$, $1$, and $2$.
In contrast to the $d^1$ case, the Heisenberg exchange changes sign from AFM ($J>0$) to FM ($J<0$) as the Hund's coupling increases.
The relative importance of the $\kappa$ term increases as $t_{a_{1g}}/t$ decreases.
Although $\kappa$ is negative in this case, its sign can be reversed by a global rotation of the orbital pseudospins, $\vb*{\tau}_i \to -\vb*{\tau}_i$, without changing the Heisenberg exchange.
The resulting exchange parameters therefore provide access to the regime considered in the main text.
We note, however, that the $d^2$ configuration carries a local spin $S=1$, whereas the model analyzed in the main text has $S=1/2$.
Since the Hamiltonian contains no spin-quadrupolar interactions, the qualitative features of the LFW analysis are expected to remain similar, although quantitative results, such as the critical coupling strengths, may depend on $S$.

\subsection{Hopping integrals in edge-sharing octahedra}

We finally examine whether the hopping hierarchy required for the minimal spin-orbital model can arise in edge-sharing octahedra. 
We consider neighboring transition-metal ions surrounded by octahedra with parallel trigonal axes and evaluate the hopping integrals using the Slater-Koster parametrization.

The trigonal distortion is characterized by the angle $\vartheta$ between a metal-ligand bond and the threefold axis.
For an undistorted octahedron, $\vartheta_0=\cos^{-1}(1/\sqrt{3})\simeq54.7^\circ$.
Trigonal elongation and compression correspond to $\vartheta<\vartheta_0$ and $\vartheta>\vartheta_0$, respectively.

The hopping between neighboring transition-metal ions consists of direct and ligand-mediated contributions. 
To investigate their relative importance, we introduce an interpolation parameter $\xi$:
\begin{equation}
T(\xi) = \xi\, T_{\rm dir} +(1-\xi)\, T_{\rm lig},
\end{equation}
where $T_{\rm dir}$ and $T_{\rm lig}$ denote the direct and ligand-mediated hopping matrices, respectively. 
The limits $\xi=1$ and $\xi=0$ correspond to purely direct and purely ligand-mediated hopping.

We use the Slater-Koster parameters $t_{dd\sigma}$, $t_{dd\pi}$, and $t_{dd\delta}$ for direct hopping and $t_{pd\sigma}$ and $t_{pd\pi}$ for metal-ligand hopping~\cite{Slater1954}. 
The charge-transfer energy is denoted by $\Delta_p$. 
Throughout the analysis, we set $t_{pd\pi}^2/\Delta_p=0.40$~meV and $t_{dd\sigma} = -0.40$~meV, and adopt the ratios $t_{pd\sigma} : t_{pd\pi} = -\sqrt{3}:1$ and $t_{dd\sigma} : t_{dd\pi} : t_{dd\delta} = 6:-4:1$. 

For each pair $(\xi,\vartheta)$, we transform the hopping matrix into the basis $\{\ket*{z,+},\ket*{z,-},\ket*{0}\}$.
To quantify the suppression of hopping within the $e_g^\pi$ doublet relative to the hopping between the doublet and the $a_{1g}$ singlet, we define
\begin{equation}
r = \max\left(\left|\frac{t_+}{t}\right|,\left|\frac{t_-}{t}\right|\right),
\end{equation}
where $t_\pm$ denote the hopping integrals between neighboring $\ket*{z,\pm}$ orbitals, respectively. 
A small value of $r$ indicates that the intra-doublet hopping is suppressed relative to the inter-multiplet hopping responsible for the $\kappa$ term.

\begin{figure*}[t]
	\centering
	\includegraphics[width=14cm]{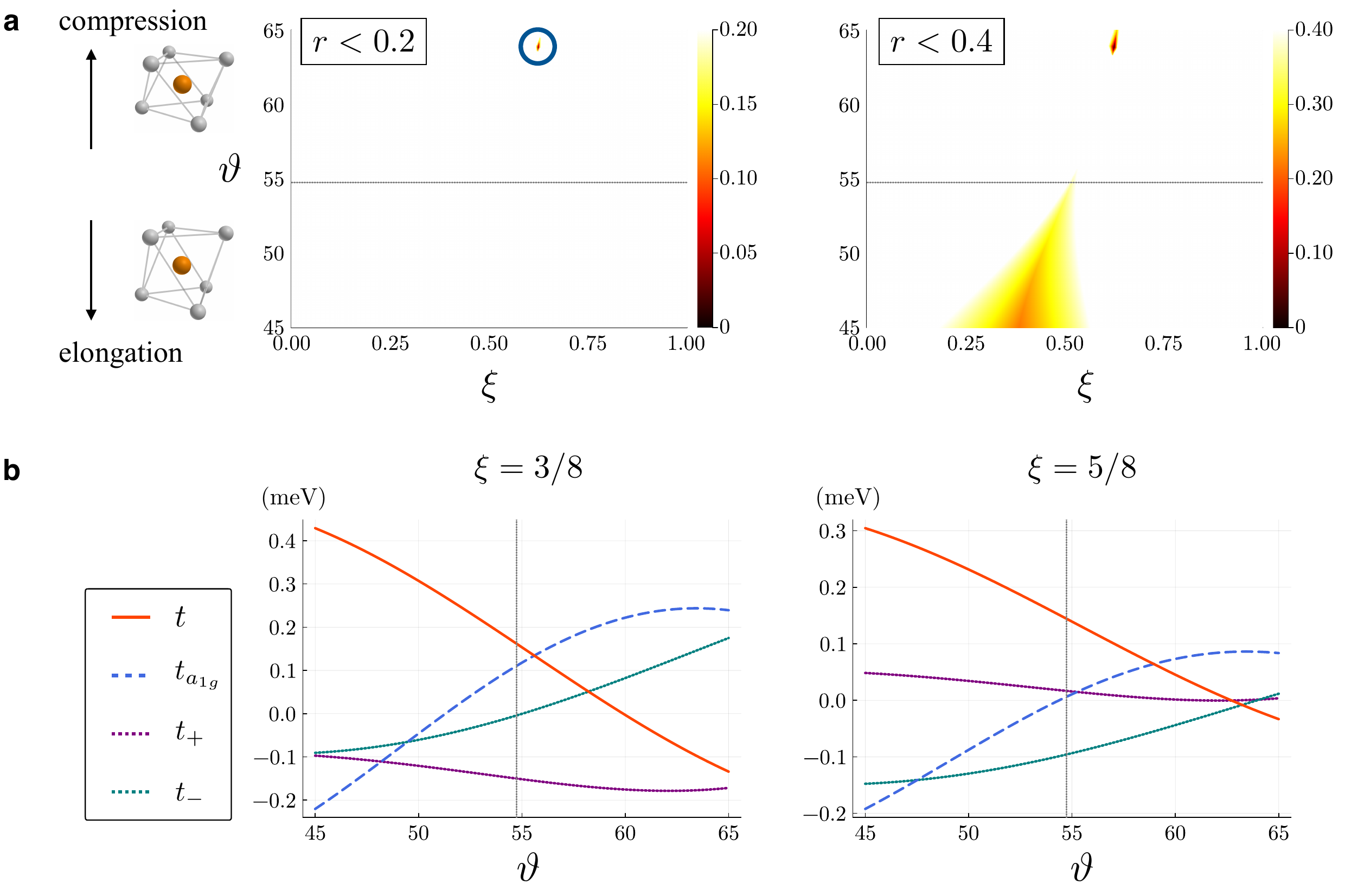}
	\caption{
		\textbf{Hopping integrals in edge-sharing octahedra.}
		\textbf{a} Color maps showing the regions satisfying $r<0.2$ (left) and $r<0.4$ (right) in the $(\xi,\vartheta)$ plane. The colored regions satisfy the respective criteria, with darker colors indicating smaller values of $r$. The region satisfying $r<0.2$ is highlighted by a circle.
		\textbf{b} Hopping integrals $t$, $t_{a_{1g}}$, $t_+$, and $t_-$ as functions of the trigonal distortion angle $\vartheta$ for $\xi=3/8$ (left) and $5/8$ (right).
		The dashed gray lines indicate the undistorted-octahedron angle $\vartheta_0=\cos^{-1}(1/\sqrt{3})\simeq54.7^\circ$.
	}
	\label{fig:supple_hopping}
\end{figure*}

Figure~\ref{fig:supple_hopping}(a) shows the regions in the $(\xi,\vartheta)$ plane satisfying $r<0.2$ and $r<0.4$.
The stronger suppression criterion, $r<0.2$, is satisfied only in a narrow region on the trigonal-compression side, $\vartheta>\vartheta_0$, with no corresponding region on the elongation side, $\vartheta<\vartheta_0$.
For the less restrictive criterion, $r<0.4$, an extended region also appears on the elongation side.

Figure~\ref{fig:supple_hopping}(b) shows the hopping integrals as functions of $\vartheta$ for two representative values of $\xi$. 
For $\xi=3/8$, the intra-doublet hopping is moderately suppressed around $\vartheta\simeq45^\circ$-$50^\circ$, corresponding to trigonal elongation. 
However, other hopping channels remain appreciable. 
For $\xi=5/8$, a stronger suppression occurs near $\vartheta\simeq64^\circ$ on the compression side, where both intra-doublet hopping integrals become small compared with the inter-multiplet hopping. 
In this regime, the hopping between $a_{1g}$ orbitals remains substantial.

These results demonstrate that the hopping hierarchy underlying the minimal spin-orbital model can arise in edge-sharing octahedra. 
However, the region exhibiting strong suppression of intra-doublet hopping is relatively narrow, indicating that a realization dominated by the $J$ and $\kappa$ interactions may require a suitable balance between direct and ligand-mediated hopping processes.

\section{Effects of additional exchange interactions on the dimer configurations}

In this section, we examine how additional orbital interactions and spin-orbit coupling (SOC) affect the subextensively degenerate dimer configurations of the minimal spin-orbital model. 
For each interaction, we first analyze whether it lifts the degeneracy within the dimer configurations and then examine the stability of the dimer configuration using site-product variational calculations.
In the variational calculations, we set $J=\kappa=1$ and $\varrho_\parallel=\varrho_\perp=0.8$, and introduce each additional interaction separately, varying its coupling constant from $-1$ to $1$. 

We first consider the bilinear orbital exchanges, which can be rewritten as
\begin{equation}
\mathcal{H} = \sum_{\langle i,j \rangle} \Big[ 
K_{zz} (1+\varrho_{\parallel} \vb*{S}_i\cdot\vb*{S}_j) \tau_i^z \tau_j^z
+(1+\varrho_{\perp}\vb*{S}_i\cdot\vb*{S}_j) \Big( 4K_{\pm\pm} \tilde{\tau}_i^\gamma \tilde{\tau}_j^\gamma 
- 2(K_{\pm}+K_{\pm\pm}) \vb*{\tau}_i^{\rm pl}\cdot\vb*{\tau}_j^{\rm pl}\Big)\Big], 
\end{equation}
where $\vb*{\tau}_i^{\rm pl}=(\tau_i^x,\tau_i^y)$.
These interactions consist of an orbital dipolar term, a bond-dependent Ising (compass) term, and an orbital XY term. 
We examine the three contributions separately. 

\subsection{Orbital dipolar interaction}

We first consider the interaction proportional to $K_{zz}$.
Within the dimer configurations, the orbital pseudospins lie in the $xy$-plane, satisfying $\ev*{\tau_i^z}=0$. 
Since the local spin and orbital states are separable, the expectation value of this interaction vanishes for every dimer configuration. 
Consequently, the orbital dipolar interaction does not lift the subextensive degeneracy. 

We next examine whether this interaction destabilizes the dimer configurations. 
The variational calculations show that the dimer configurations remain stable over a substantial range of $K_{zz}$, with transitions to competing states occurring at approximately $K_{zz}=-0.82$ and $0.84$. 

\subsection{Orbital compass interaction}

We next consider the bond-dependent Ising interaction proportional to $4K_{\pm\pm}$.
To examine its effect on the subextensive degeneracy, we compare two neighboring $xy$-chains with either identical or different dimer orientations, corresponding to locally uniform and alternating arrangements, respectively. 
Since the contributions from the $x$- and $y$-bonds are identical for the two arrangements, we need only compare the contributions from the $z$-bond. 
For a $z$-bond, the orbital pseudospin projections satisfy $\ev*{\tilde{\tau}_i^z}=\ev*{\tilde{\tau}_j^z}=1/4$ in both arrangements. 
The corresponding compass interaction therefore contributes equally to their energies and does not lift the degeneracy among the dimer configurations. 

The variational calculations further show that the dimer configurations remain stable over a substantial range of the compass coupling, with a transition to a competing state occurring at approximately $4K_{\pm\pm}=0.63$. 
No transition is found on the negative-coupling side within the investigated range. 

\subsection{Orbital XY interaction}

Unlike the preceding interactions, the orbital XY interaction lifts the subextensive degeneracy by favoring a particular arrangement of neighboring dimer chains. 
To demonstrate this, we again compare the locally uniform and alternating arrangements. 
For the $z$-bond connecting two neighboring chains, their energy contributions are 
\begin{equation}
E_{\rm uni}^{(z)} =(1-\varrho_\perp S^2) \left[ -\frac{1}{2} (K_\pm +K_{\pm\pm}) \right], \quad
E_{\rm alt}^{(z)} =(1-\varrho_\perp S^2) \left[ +\frac{1}{4} (K_\pm +K_{\pm\pm}) \right], 
\end{equation}
respectively. 
For $1-\varrho_\perp S^2>0$, the uniform-dimer configuration is favored when $K_\pm+K_{\pm\pm}>0$, whereas the alternating-dimer configuration is favored when $K_\pm+K_{\pm\pm}<0$.

The variational calculations confirm this selection mechanism. 
Writing the coefficient of the orbital XY interaction as $K_{\rm XY}=-2(K_\pm+K_{\pm\pm})$, we find that negative $K_{\rm XY}$ favors the uniform-dimer configuration, while positive $K_{\rm XY}$ favors the alternating-dimer configuration. 
The uniform-dimer configuration remains stable throughout the investigated range, whereas the alternating-dimer configuration becomes unstable toward a competing state at approximately $K_{\rm XY}=0.30$.

\subsection{Spin-orbit coupling}

Finally, we consider the on-site SOC projected onto the $e_g^\pi$ doublet~\cite{Nakai2026},
\begin{equation}
\mathcal{H}_{\rm SOC} = \lambda_\parallel \sum_i \tau_i^z S_i^z.
\end{equation}
For the dimer configurations, the orbital pseudospins lie in the $xy$-plane, and the local spin and orbital states are separable.
Consequently, $\ev*{\mathcal{H}_{\rm SOC}}=0$ for every configuration, indicating that SOC does not lift the subextensive degeneracy. 

The variational calculations reveal that the orbital pseudospins respond continuously to SOC without changing their in-plane orientations.
The spin configuration remains unchanged, while the magnitude of the in-plane orbital component decreases and a finite out-of-plane component $\ev*{\tau_i^z}$ develops.
The orbital pseudospins tilt by approximately $26.6^\circ$ away from the $xy$ plane at $|\lambda_\parallel|=1$.
Despite this orbital canting, the subextensive degeneracy remains intact, and no transition to a competing phase is found within the investigated range $-1\leq\lambda_\parallel\leq1$. 

Overall, these results demonstrate the robustness of the dimer configurations against additional interactions of appreciable strength. 
The orbital dipolar and compass interactions preserve the subextensive degeneracy but can favor competing states at sufficiently strong coupling. 
The orbital XY interaction selects between uniform and alternating dimer configurations, while the selected configurations remain stable against weak perturbations. 
SOC induces continuous orbital canting without lifting the degeneracy or favoring competing states over the investigated coupling range. 

\bibliographystyle{apsrev4-2}
\bibliography{refs}